\documentclass[a4paper,12pt]{scrartcl}
\usepackage[]{graphicx}\usepackage[]{color}

\usepackage{alltt}
\usepackage[left= 2.5cm,right = 2.5cm, bottom = 3.0cm,top=2cm]{geometry}
\usepackage[latin1]{inputenc} 
\usepackage{setspace}
\usepackage{amsmath}
\usepackage{amsthm}

\usepackage[flushleft]{threeparttable}

\usepackage{graphicx,psfrag,epsf}
\usepackage[markup=underlined]{changes}
\usepackage{amsmath}
\usepackage{mathtools} 
\usepackage{amsfonts}
\usepackage{amssymb}
\usepackage{graphics}
\usepackage{graphicx}
\usepackage{epstopdf}
\usepackage[nottoc]{tocbibind}

\usepackage[authoryear]{natbib}
\newcommand{\GG}[1]{}

\newlength{\minuslength}
\usepackage{subfig}

\usepackage{bm}
\usepackage{threeparttable}
\usepackage{booktabs}

\usepackage{acronym} 
\usepackage{here} 

\usepackage{titling}

\title{\Large Assessing the Effects of Monetary Shocks on Macroeconomic Stars: A SMUC-IV Framework\thanks{We would like to thank Max Breitenlechner, Joshua Chan, Martin Geiger, Domenico Giannone, Yufan Huang, Luis Uzeda, Benjamin Wong, Ren Zhang as well as participants of the 2025 \"Orebro Workshop on Macro- and Financial Econometrics for their constructive comments and valuable suggestions.}}

\author{Bowen Fu\thanks{Hunan University, Bowen Fu acknowledges the support of the National Natural Science Foundation of China (72303061), fu\_bowen@hotmail.com. } \and
	Chenghan Hou\thanks{Hunan University, Chenghan Hou, Corresponding author, chenghan.hou@hotmail.com.} \and
	Jan Pr\"user\thanks{TU Dortmund, Jan Pr\"user gratefully
		acknowledges the support of the German Research Foundation (DFG, 468814087), prueser@statistik.tu-dortmund.de.} }

\begin{document}

	\date{\today}
	\maketitle
	\begin{abstract}
%
		\begin{singlespace}
			\noindent
			This paper proposes a structural multivariate unobserved components model with external instrument (SMUC-IV) to investigate the effects of monetary policy shocks on key U.S. macroeconomic ``stars''---namely, the potential GDP growth, trend inflation, and the neutral  interest rate.
	A key feature of our approach is the use of an external instrument to identify monetary policy shocks within the multivariate unobserved components modelling framework. We develop an MCMC estimation method to facilitate posterior inference within our proposed SMUC-IV framework. 
	In addition, we propose a marginal likelihood estimator to enable model comparison across alternative specifications. Our empirical analysis shows that contractionary monetary policy shocks have  negative effects on the macroeconomic stars, highlighting the non-zero long-run effects of transitory monetary policy shocks.
		\end{singlespace}

	\end{abstract}
	\bigskip
	
		 \bigskip

	\thispagestyle{empty} 
	\newpage
	
	
	\pagenumbering{arabic}

\section{Introduction}
\setcounter{page}{1}


Long-run equilibrium levels of key macroeconomic variables---commonly referred to as ``stars"---are central for guiding countercyclical policy and assessing long-run performance. In the conventional view, the macroeconomic stars are unaffected by monetary policy shocks. These shocks are typically believed to have only transitory effects. However, recent evidence challenges this view. For instance, \cite{jorda2024long} show that contractionary monetary policy shocks have negative and long-lasting effects on output. A less studied empirical question, however, is how monetary policy shocks affect the macroeconomic stars. This question is of interest because it provides guidance on how policymakers can bring inflation back to target while supporting sustainable economic growth in the wake of the heightened inflation following the COVID-19 pandemic.



 The main objective of this paper is to develop a unified econometric framework to address this question.  Multivariate unobserved components models have become the standard econometric tool for analyzing macroeconomic stars \citep[e.g.,][]{kuttner1994estimating,laubach2003measuring,chan2016bounded,zaman2025unified}. These models decompose observed macroeconomic series into long-run trends and short-run cycles, identifying the stars with the trend components.  The traditional multivariate unobserved components model framework, albeit useful
to model the macroeconomic stars,  lacks a formal strategy to identify the  effects of monetary policy shocks on the stars.



  This paper contributes to the existing literature by introducing a structural multivariate unobserved components model to assess the effects of monetary policy shocks on the macroeconomic stars. The main novelty of our approach is the use of an external instrument to identify monetary policy shocks for structural analysis within a multivariate unobserved components model, which we refer to as SMUC-IV. While the use of external instruments to identify macroeconomic shocks in structural vector autoregressions has become increasingly common \citep{mertens2013dynamic,ramey2016macroeconomic,stock2018identification,jarocinski2020deconstructing,caldara2019monetary,arias2021inference}, very few studies have considered their application within multivariate unobserved components models. The SMUC-IV introduced in this paper addresses this gap.  Although the SMUC-IV developed in this paper is used to assess the effects of monetary policy shocks on the macroeconomic stars, the methodology readily extends to  broader applications in a straightforward manner.


Our proposed SMUC-IV possesses two desirable characteristics. First, it offers a unified framework that jointly estimates the macroeconomic stars and assesses their responses to monetary policy shocks, rather than focusing on a single star.  This ensures internally consistent estimates of the macroeconomic stars, enabling a coherent analysis of the dynamic interactions between the long-run trends and short-run cyclical fluctuations. Second, our SMUC-IV specification allows for correlations among the innovations of all included trend and cycle components. This flexible correlation structure has been shown to be empirically important in many studies on unobserved components models \citep{morley2003beveridge,basistha2007new,grant2017bayesian,grant2017reconciling,hwu2019estimating}.

A further contribution of this paper is the development of an MCMC estimation procedure and a marginal likelihood estimator, designed to facilitate estimation and model comparison across various variants of the proposed SMUC-IV. Specifically, our proposed MCMC method builds upon the precision sampling algorithm for Gaussian state space models developed by \cite{chan2009efficient}. A novel feature of our approach is that it exploits the joint Gaussianity between the observed data and the state parameters, which allows for the direct derivation of the conditional posterior of the states. This approach bypasses the need for separate computations of the conditional likelihood and prior, leading to a more straightforward implementation, particularly when the state and measurement equations are correlated \citep{grant2017bayesian, grant2017reconciling, leiva2023endogenous}. A related benefit of recognizing the joint Gaussianity between the observed data and the state parameters is that it permits direct derivation of the likelihood function unconditional on the high-dimensional latent states. This unconditional likelihood is an essential component for our proposed marginal likelihood estimator. Specifically, we construct a conditional Monte Carlo improved modified harmonic mean estimator to compute the marginal likelihoods for various specifications of the SMUC-IV. The conditional Monte Carlo method was recently proposed by \citet{chan2023comparing} for estimating marginal likelihood for large vector autoregressions, and has been shown to significantly improve estimation accuracy.

In our empirical analysis, we jointly estimate three US macroeconomic stars---the  potential GDP growth, trend inflation, and the neutral  interest rate. For the identification of the monetary policy shock, we employ the orthogonal high-frequency surprise series of \citet{bauer2023reassessment} as our external instrument for monetary policy shocks. The series expands the set of events to include Federal Reserve Chair speeches and orthogonalizes the surprises to pre-announcement variables to strengthen relevance and exogeneity. 

To assess the validity of our framework, we conduct a Bayesian model comparison exercise in our empirical analysis. The results show that the proposed SMUC-IV is more strongly supported by the data than competing specifications. The findings support the relevance of the external instrument for monetary policy shocks, confirm correlations between long-run trends and short-run cycles, and---most importantly---underscore the influence of monetary policy shocks on the macroeconomic stars.

 We document two main findings on how monetary policy shocks affect the macroeconomic stars.  First, contractionary monetary policy shocks lead to declines in the macroeconomic stars. The negative effects on potential GDP growth are consistent with the empirical evidence in \cite{moran2018innovation}, \cite{garga2021output}, \cite{ma2023monetary}, \cite{jorda2024long}, and \cite{meier2024monetary}, which shows that tighter monetary policy dampens investment in innovation and, in turn, productivity. Likewise, the decline in the neutral interest rate following contractionary monetary policy shocks may be interpreted as reflecting the same adverse effects of monetary tightening on investment in innovation. The fall in trend inflation is consistent with findings that contractionary monetary policy shocks lower long-term inflation expectations \citep[e.g.,][]{jarocinski2020deconstructing,diegel2021long}. Second, counterfactual analysis based on historical decompositions shows that, in the absence of these contractionary shocks, potential GDP growth, trend inflation, and the neutral interest rate would have been notably higher, implying that monetary policy is a driving force in shaping the stars. Our two main findings are robust across a range of alternative specifications.
 
The negative effects of contractionary monetary policy shocks on potential GDP growth and the neutral interest rate can be rationalized by the mechanism highlighted in the Keynesian endogenous growth model with nominal rigidities in \cite{fornaro2023scars}, which shows that monetary tightening may trigger a ``supply--demand doom loop": lower demand reduces firms' profits, which curtails investment; weaker investment slows expected productivity growth, lowers household wealth, and further depresses demand---ultimately leading to substantial declines in potential GDP growth and the neutral  interest rate. The fall in trend inflation may reflect a re-anchoring channel, whereby expectations that have drifted away from the inflation target are deliberately pulled back, as discussed in detail in \cite{diegel2021long}. The results imply that central banks face a dilemma: disinflating the economy may come at the cost of lower potential GDP growth and a lower neutral interest rate. This, in turn, points to a potential role for fiscal interventions that support business investment and the economy's productive capacity during periods of disinflation.

The paper is organized as follows. Section 2 introduces the proposed SMUC-IV. Section 3 details the prior distributions and develops an efficient MCMC method for estimation. Section 4 presents our modified harmonic mean estimator for marginal likelihood estimation, improved via Monte Carlo methods. Section 5 describes the data and empirical results, and Section 6 concludes.

\section{Econometric Framework}
In this section, we introduce our SMUC-IV. Section~\ref{sec:Model Specification} presents the specification of the structural multivariate unobserved components model. Section~\ref{sec:Identification via Proxy Variable} then outlines how the external instrument is employed for structural identification.


\subsection{Model Specification}\label{sec:Model Specification}
We consider the following trend and cycle decomposition:
\begin{align}
	g_{t} &= g_{t}^{*} + c_{g,t}, \label{eq:g}  \\ 
	\pi_{t} &= \pi_{t}^{*} + c_{\pi,t}, \\
	i_{t} &= \pi_{t}^{*} + r_{t}^{*} + c_{i,t}, \label{eq:r}
\end{align}
where $g_t$ denotes real GDP growth, $\pi_t$ denotes inflation, and $i_t$ denotes the nominal interest rate. In addition, $g_{t}^{*}$, $\pi_{t}^{*}$, and $r_{t}^{*}$ represent the potential GDP growth, trend inflation, and the neutral interest rate, respectively. In this paper, we use the log difference of real GDP as the real GDP growth, GDP deflator inflation as the inflation, and the federal funds effective rate  as the nominal interest rate (see Section~\ref{Sec:data} for more details). We assume inflation $\pi_{t}$ and  nominal interest rate $i_{t}$ share a common trend component, $\pi_{t}^{*}$ as in \cite{del2017safety}.

We assume the trend components of real GDP growth, inflation and real interest rate to follow random walk processes:
\begin{align}
	g_t^{*} &= g_{t-1}^{*} + u_t^{g^{*}}, \label{eq:delta_g} \\
	\pi_t^{*} &= \pi_{t-1}^{*} + u_t^{\pi^{*}}, \\
	r_t^{*} &= r_{t-1}^{*} + u_t^{r^{*}}. \label{eq:rstar}
\end{align}
The initial conditions $g_0^{*}, \pi_0^{*}$  and $r_0^{*}$ are treated as parameters to be estimated. Our trends are defined, consistent with the Beveridge-Nelson decomposition \citep{beveridge1981new}, as the infinite-horizon forecast of the actual variables of interest, conditional on the information set available in period $t$, which implies a random walk for the trends and stationary, mean-zero cycles. 

The specifications of trend components broadly aligns with unobserved components models used to estimate macroeconomic stars,  including strands that focus separately on the potential GDP growth \citep[e.g.,][]{grant2017bayesian,grant2017reconciling}, trend inflation \citep[e.g.,][]{chan2013new,chan2018new,mertens2016measuring,stockwatson2,stock2016core,hwu2019estimating,eo2023understanding}, and the neutral  interest rate \citep[e.g.,][]{laubach2003measuring,holston2017measuring,del2017safety}.

Regarding the cycle components, let $\mathbf{c}_t = ( c_{g,t}, c_{\pi,t}, c_{i,t})'$ be a vector of cycle components, and we assume that $\mathbf{c}_t$ evolves according to the following VAR($p$) process:
\begin{align}
	\mathbf{c}_t = \boldsymbol{\Phi}_{1} \mathbf{c}_{t-1} + \cdots + \boldsymbol{\Phi}_{p} \mathbf{c}_{t-p} + \mathbf{u}_{t}^{c},
	\label{eq:c}
\end{align}
where $\boldsymbol{\Phi}_{1}, \ldots, \boldsymbol{\Phi}_{p}$ are $3 \times 3$ autoregressive coefficient matrices and $\mathbf{u}_t^{c}$ are the reduced-form innovations.\footnote{For simplicity, we set the initial conditions $\mathbf{c}_{0} = \ldots = \mathbf{c}_{1-p} = \mathbf{0}$.}

The state equations given in \eqref{eq:delta_g} - \eqref{eq:rstar} can be expressed more compactly. To be specific, let $\boldsymbol{\tau}_t = (g_t^*, \pi_t^*, r_t^*)'$, 
we can rewrite \eqref{eq:delta_g} - \eqref{eq:rstar} as 
\begin{align}
	\boldsymbol{\tau}_t = \boldsymbol{\tau}_{t-1} + \mathbf{u}_t^{\tau}, \label{eq:tau}
\end{align}
where $\mathbf{u}_t^{\tau} = (u_t^{g^*}, u_t^{\pi^*}, u_t^{r^*} )'$. Stacking equation~\eqref{eq:tau} over equation~\eqref{eq:c}, our model can be represented as
\begin{align}
	\boldsymbol{\eta}_t = \mathbf{A}_1 \boldsymbol{\eta}_{t-1} + \cdots + \mathbf{A}_p \boldsymbol{\eta}_{t-p} + \mathbf{u}_t, \label{eq:SVAR}
\end{align}
where $\boldsymbol{\eta}_t = (\boldsymbol{\tau}_t', \mathbf{c}_t')'$ is a vector containing the trend and cycle components. The coefficient matrices are $\mathbf{A}_1 = \text{diag}(\mathbf{I}_3,\boldsymbol{\Phi}_1)$, $\mathbf{A}_i = \text{diag}(\mathbf{0}_{3 \times 3},\boldsymbol{\Phi}_i)$ for $i = 2, \ldots, p$. The residual vector $\mathbf{u}_t = (\mathbf{u}_t^{\tau'}, \mathbf{u}_t^{c'} )$ is of dimension $6 \times 1$, which will be described shortly. To complete our model specification, we assume that the residual vector $\mathbf{u}_t$ is related to the structural shocks by
\begin{align}
	\mathbf{u}_t = \mathbf{B} \boldsymbol{\epsilon}_t, \quad \boldsymbol{\epsilon}_t \sim \mathcal{N}(\mathbf{0}_{6 \times 1}, \mathbf{I}_{6}), \label{eq:Be}
\end{align}
where  $\boldsymbol{\epsilon}_t$ is a vector of structural shocks, $\mathbf{B}$ is the contemporaneous response matrix that is assumed to be non-singular, and $\boldsymbol{\Sigma} = \mathbf{B} \mathbf{B}'$ is the covariance matrix of the residual vector $\mathbf{u}_t$. It is known that the contemporaneous response matrix $\mathbf{B}$, hence, the structural shocks $\boldsymbol{\epsilon}_t$, cannot be separately identified without additional information.  In this paper, we employ an external instrument to identify monetary policy shocks.  The next section provides the details of our identification approach.

Our proposed model specified in equations \eqref{eq:SVAR}--\eqref{eq:Be} provides a unified framework for studying the macroeconomic stars and their corresponding cycle components. Furthermore, we allow the innovations of all trend and cycle components, i.e., $\mathbf{u}_t$, to be correlated.  This contrasts with conventional studies on unobserved components models \citep[e.g.,][]{watson1986univariate,stockwatson2,chan2018new,laubach2003measuring,zaman2025unified}, which typically assume these innovations to be independent. Specifically, the contemporaneous response matrix $\mathbf{B}$ is assumed to be non-singular and unrestricted, which implies that  $\boldsymbol{\Sigma} = \mathbf{B} \mathbf{B}'$ is a full covariance matrix.  In our empirical analysis, we assess the validity of this full correlation structure through a formal Bayesian model comparison exercise, which indicates strong evidence in favor of this modeling feature supported by the data.

\subsection{Identification via External Instrument}\label{sec:Identification via Proxy Variable}


In this section, we first discuss how the  external instrument can be incorporated into our modelling framework for identification, and then describe our proposed SMUC-IV as an augmented structural vector autoregression, a form which is used later for Bayesian estimation. In this paper, we consider the case in which one external instrument is used to identify one structural shock, namely, the monetary policy shock of interest. For the case of using multiple external instruments to identify multiple structural shocks, we refer readers to \cite{arias2021inference}, \cite{braun2023identification}, and \cite{hou2024large} for more details.

To set the stage, we designate the last shock in $\boldsymbol{\epsilon}_t$, denoted by $\epsilon_{m,t}$, as the monetary policy shock of interest. Accordingly, we can write the structural shocks as $\boldsymbol{\epsilon}_t = (\boldsymbol{\epsilon}_{-m,t}', \epsilon_{m,t}')'$  where $\boldsymbol{\epsilon}_{-m,t}$ is a vector containing all structural shocks other than $\epsilon_{m,t}$. Suppose that an external instrument $m_t$ is available, which is linked to the structural shocks $\boldsymbol{\epsilon}_t$ as
\begin{align}
	m_t = \boldsymbol{\gamma}' \boldsymbol{\epsilon}_t + \alpha v_t, \quad v_t \sim \mathcal{N}(0,1), \label{eq:proxy}
\end{align} 
where $\boldsymbol{\gamma}$ is a $6 \times 1$ vector of coefficient parameters associated with the structural shocks $\boldsymbol{\epsilon}_t$ and $v_t$ is the shock of the external instrument equation independent of $\boldsymbol{\epsilon}_t$, which can be interpreted as the measurement error of the  external instrument.

To achieve identification of the monetary policy shock $\epsilon_{m,t}$, the external instrument $m_t$ is required to be correlated with $\epsilon_{m,t}$, but uncorrelated with $\boldsymbol{\epsilon}_{-m,t}$. More precisely, a valid external instrument needs to satisfy the following relevance and exogeneity conditions:  
\begin{align*}
	&\text{Relevance condition}: E(m_t \epsilon_{m,t}) = \beta \neq 0, \\
	&\text{Exogeneity condition}: E(m_t \boldsymbol{\epsilon}_{-m,t}') = \mathbf{0}_{1 \times 5}.
\end{align*}
These two conditions are central to understanding how the external instrument can be used to identify the structural shock of interest $\epsilon_{m,t}$. Specifically, it conveys identifying information by distinguishing $\epsilon_{m,t}$ from the other shocks in $\boldsymbol{\epsilon}_{-m,t}$ through differences in their covariance structures with  the external instrument $m_t$. 

The relevance and exogeneity conditions together provide further information by imposing zero restrictions on the parameter vector $\boldsymbol{\gamma}$. To see this, we first compute the covariance between $m_t$ and $\boldsymbol{\epsilon}_t$, that is
\begin{align*}
	E(m_t \boldsymbol{\epsilon}_t') = \left( E(m_t \boldsymbol{\epsilon}_{-m,t}'),  E(m_t \epsilon_{m,t}) \right) =  (\mathbf{0}_{1 \times 5}, \beta)',
\end{align*}
where the last equality is implied by the relevance and exogeneity conditions. On the other hand, given the  external instrument equation \eqref{eq:proxy}, the covariance between $m_t$ and $\boldsymbol{\epsilon}_t$ can also be expressed as
\begin{align*}
	E(m_t \boldsymbol{\epsilon}_t') = E \left( (\boldsymbol{\gamma}' \boldsymbol{\epsilon}_t  + \alpha v_t )\boldsymbol{\epsilon}_t' \right) = \boldsymbol{\gamma}',
\end{align*}
where the second equality holds because $v_t$ and $\boldsymbol{\epsilon}_t$ are assumed to be uncorrelated. Therefore, these results imply that
\begin{align}
	\boldsymbol{\gamma} = (\mathbf{0}_{1 \times 5}, \beta)'. \label{eq:zero_gamma}
\end{align}
To summarize, our SMUC-IV is specified as equations \eqref{eq:SVAR}, \eqref{eq:Be}, and \eqref{eq:proxy}, subject to the zero restrictions given in \eqref{eq:zero_gamma}. The external instrument equation~\eqref{eq:proxy} with the zero restrictions in \eqref{eq:zero_gamma} gives $m_t = \beta \epsilon_{m,t} + \alpha v_t$, which is consistent with the specification used in \cite{caldara2019monetary}.

Our proposed SMUC-IV can be represented more compactly as an augmented structural vector autoregression for $\widetilde{\boldsymbol{\eta}}_t = (\boldsymbol{\eta}_t', m_t)'$:
\begin{align}
	\widetilde{\boldsymbol{\eta}}_t = \widetilde{\mathbf{A}}_1 \widetilde{\boldsymbol{\eta}}_{t-1} + \cdots + \widetilde{\mathbf{A}}_p \widetilde{\boldsymbol{\eta}}_{t-p} + \widetilde{\mathbf{B}} \widetilde{\boldsymbol{\epsilon}}_t, \quad \widetilde{\boldsymbol{\epsilon}}_t \sim \mathcal{N}(\mathbf{0}_{7 \times 1}, \mathbf{I}_{7}), \label{eq:aug_SVAR}
\end{align} 
where $\widetilde{\boldsymbol{\epsilon}}_t = (\boldsymbol{\epsilon}_t', v_t)'$. The parameter matrices are given by
\begin{align}
	\widetilde{\mathbf{A}}_i &= 
	\begin{pmatrix}
		\mathbf{A}_i & \mathbf{0}_{6 \times 1} \\
		\mathbf{0}_{1 \times 6} & 0
	\end{pmatrix} \text{ for $i = 1, \ldots, p$}, \label{eq:zero_A}
	\\
	\widetilde{\mathbf{B}} &= 
	\begin{pmatrix}
		\mathbf{B} & \mathbf{0}_{6 \times 1} \\
		\boldsymbol{\gamma}' & \alpha
	\end{pmatrix} \text{ with $\boldsymbol{\gamma} =  (\mathbf{0}_{1 \times 5}, \beta)'$}. \label{eq:zero_B}
\end{align}	
This augmented structural vector autoregression representation is commonly used in Bayesian analysis to facilitate posterior inference.

As indicated by recent studies \citep{arias2021inference,braun2023identification,hou2024large}, the use of  external instruments can only achieve set-identification. For instance, in our case, it can be shown that the second-to-last column of $\widetilde{\mathbf{B}}$, which embeds the impact responses to the monetary policy shock and corresponds to the last column of $\mathbf{B}$, can only be identified up to a sign change. A common solution to this set-identification issue is to impose sign restrictions on the impulse responses, which are informed by economic theory. In this paper, rather than imposing dogmatic restrictions on response directions, we identify monetary policy shocks using an informative prior that accommodates estimation uncertainty. We will discuss the prior in more detail in Section~\ref{sec:Priors}.

\section{Bayesian Estimation}\label{sec:Bayesian Estimation}

\subsection{Priors}\label{sec:Priors}
This section describes the prior distributions assigned to the model parameters. Let $\bar{\boldsymbol{\tau}}_0 = (g_0^{*}, \pi_0^{*}, r_0^{*})'$ denote the vector of the initial state parameters, $\boldsymbol{\Phi}_{l,i,j}$ the $(i,j)$th element of the autoregressive coefficient matrix $\boldsymbol{\Phi}_{l}$ for $i,j = 1, 2, 3$, $l = 1, \ldots, p$, and $\mathbf{B}_{i,j}$ the $(i,j)$th element of the contemporaneous impact matrix $\mathbf{B}$ for $i,j = 1, \ldots, 6$. We assume the following independent priors:
\begin{align}
	&\boldsymbol{\Phi}_{l,i,j} \sim \mathcal{N}( \phi_{l,i,j}, V_{\phi,l,i,j}), 
	\quad 
	\mathbf{B}_{i,j} \sim \mathcal{N}(b_{i,j}, V_b), 
	\quad 
	\bar{\boldsymbol{\tau}}_0 \sim \mathcal{N}(\bar{\boldsymbol{\tau}}_{00}, \mathbf{V}_{\bar{\tau}_{00}}), \nonumber \\
	&\beta \sim \mathcal{N}(\beta_{0}, V_{\beta}), 
	\quad 
	\alpha \sim \mathcal{N}(\alpha_0, V_{\alpha}) \mathbf{1}(\alpha > 0).
	\label{eq:prior}
\end{align}

Moreover, we consider a Minnesota-type adaptive hierarchical shrinkage prior for the autoregressive coefficients to address overfitting concerns in our richly parametrized model. To be specific, we set the prior mean of the autoregressive coefficient to $\phi_{l,i,j} = 0$ and the prior variance to 
\begin{align*}
	\mathbf{V}_{\phi,l,i,j} = 
	\begin{dcases}
		\frac{ \kappa_1 }{ l^{2} }, \quad \text{$i = j$, $i, j = 1, 2, 3$, $l = 1, \ldots, p$}, \\
		\frac{ \kappa_2 \sigma^2_i }{l^2 \sigma^2_j}, \quad \text{$i \neq j$, $i,j = 1, 2, 3$, $l = 1, \ldots, p$}.
	\end{dcases}
\end{align*}
 This specification reflects the prior belief that the coefficients on more distant lags are less important than those on recent lags, and are therefore shrunk more strongly toward zero. Following standard practice, the scale parameter $\sigma^2_i$ is set equal to the residual variance of an AR($p$) model for its corresponding variable $i$. The hyperparameters $\kappa_1$ and $\kappa_2$ control the shrinkage strength for the own-lag and cross-variable-lag coefficients, respectively. Empirical evidence from recent studies, such as \cite{cross2020macroeconomic} and \cite{chan2021minnesota}, indicates that allowing the hyperparameters of shrinkage prior to be parameters to be estimated can substantially improves both forecasting accuracy and model fit. Therefore, rather than fixing these hyperparameters at predetermined values, we treat $\kappa_1$ and $\kappa_2$ as unknown parameters and assign them the following prior distributions:
\begin{align*}
	\kappa_1 \sim \mathcal{U}(0,1), \quad \kappa_2 \sim \mathcal{U}(0,1).
\end{align*}

We use an informative prior for the contemporaneous impact matrix $\mathbf{B}$, setting the prior variance to $V_{b} = 0.01$ and specifying the prior mean as follows:\footnote{In Section~\ref{sec:Robustness checks}, we conduct a robustness analysis by treating $V_b$ as a unknown parameter to be estimated and the main finding our empirical results remain unchanged.}
\begin{align*}
	b_{i,j} = 
	\begin{dcases}
		0.1, \quad \text{$i = j$, $i, j = 1, 2, 3$}, \\
		1, \quad \text{$i = j$, $i,j = 4,5,6$}, \\
		0, \quad \text{$i \neq j$, $i,j = 1, \ldots, 6$.} 
	\end{dcases}
\end{align*}
This prior centers the contemporaneous impact matrix $\mathbf{B}$ on a diagonal matrix, implying a priori uncorrelatedness among all reduced-form residuals for the cycle and trend components, with the prior standard deviations set to $0.1$ for  the trend components and $1$ for the cycle components.\footnote{Informative priors are commonly used in the estimation of multivariate unobserved component models. By imposing small standard deviations on the latent state parameters, this prior belief helps prevent overfitting and yields smoother, more economically sensible estimates.} 

The prior we consider here provides additional information for identifying the contemporaneous response matrix $\mathbf{B}$. Recall that the monetary policy shock $\epsilon_{m,t}$ is ordered last in $\boldsymbol{\epsilon}_t$. That means, the last column of $\mathbf{B}$ is the impact responses of the trend and cycle components to a monetary policy shock. Our prior reflects a belief that a one-standard-deviation increase in monetary policy shock $\epsilon_{m,t}$ results in a $100$ basis point increase in the cycle of nominal interest rate $c_{i,t}$, while having no effect on $g_t^*$, $\pi_t^*$, $r_t^*$, $c_{g,t}$ and $c_{\pi,t}$, on impact. By the definition of $c_{i,t}$ in \eqref{eq:r}, the impact response of $c_{i,t}$ to a monetary policy shock is entirely attributable to the increase in $i_t$. This identifying information from our prior is similar to conventional studies on identifying monetary policy shock by normalize the magnitude of the nominal interest rate response to a positive value.\footnote{For instance, \cite{bauer2023reassessment} normalize the nominal interest rate response to a monetary policy shock by $25$ basis points, while \cite{miranda2021transmission} use a normalization of $100$ basis points. Unlike these approaches, we do not fix the size of the response of nominal interest rate to a monetary policy shock, instead, we impose a relatively tight prior that centers the response at $100$ basis points.} We also highlight that since we use an informative prior that centers the impact responses of $g_t^*$, $\pi_t^*$, $r_t^*$, $c_{g,t}$ and $c_{\pi,t}$ to a monetary policy shock at zeros, any nonzero effects found in our empirical study must be supported by the data.


For the parameters in the external instrument equation, we set $V_{\beta} = 1$, $V_{\alpha} = 1$, $\alpha_0 = 0$, and $\beta_0 = 0.5 \times \sigma_m$, where $\sigma_m$ denotes the standard deviation of the external instrument. This is comparable to the setting in \cite{caldara2019monetary}. For the initial state parameters $\bar{\boldsymbol{\tau}}_0$, we use an uninformative prior by setting $\mathbf{V}_{\bar{\tau}_{00}} = 100 \mathbf{I}_3$ and $\bar{\boldsymbol{\tau}}_{00} = (g_1, \pi_1, i_1)'$ where  $g_1$, $\pi_1$ and $i_1$ are the first observations of log real GDP, inflation, and the nominal interest rate in the sample period, respectively.

\subsection{Posterior Sampler}\label{sec:Posterior Sampler}
We now discuss the estimation of our model with the prior described in the previous section. To set the stage, let $\mathbf{y} = (\mathbf{y}_1', \ldots, \mathbf{y}_{T}')'$, where $\mathbf{y}_t = ( g_t, \pi_t, i_t, m_t)'$, denote the vector of observed data, $\boldsymbol{\Phi} = (\boldsymbol{\Phi}_1, \ldots, \boldsymbol{\Phi}_p)$ denote the collection of the autoregressive coefficient matrices and $\boldsymbol{\tau} = (\bar{\boldsymbol{\tau}}_0', \boldsymbol{\tau}_1', \ldots, \boldsymbol{\tau}_T' )'$ denote a vector of state parameters. The joint posterior distribution can be simulated by sequentially sampling from the following conditional distributions:
\begin{enumerate}
	\item $p(\boldsymbol{\tau} | \mathbf{B}, \beta, \alpha, \boldsymbol{\Phi}, \kappa_{1}, \kappa_{2},\mathbf{y})$;
	\item $p( \mathbf{B}, \beta, \alpha | \boldsymbol{\tau}, \boldsymbol{\Phi}, \kappa_{1}, \kappa_{2}, \mathbf{y})$;
	\item $p( \boldsymbol{\Phi}| \boldsymbol{\tau}, \mathbf{B}, \beta, \alpha, \kappa_{1}, \kappa_{2}, \mathbf{y})$;
	\item $p(\kappa_1 | \boldsymbol{\tau}, \mathbf{B}, \beta, \alpha, \boldsymbol{\Phi}, \kappa_2, \mathbf{y})$;
	\item $p(\kappa_2 | \boldsymbol{\tau}, \mathbf{B}, \beta, \alpha, \boldsymbol{\Phi}, \kappa_1, \mathbf{y})$.
\end{enumerate}

In Step 1, we sample the state vector $\boldsymbol{\tau}$ using the precision sampling method \citep{chan2009efficient, mccausland2011simulation, rue2001fast} instead of traditional Kalman filter techniques. The precision-based sampling method has gained increasing popularity for estimating state space models due to its computational efficiency and straightforward implementation. One complication under our framework, however, involves determining the conditional posterior distribution for $\boldsymbol{\tau}$. The conventional approach requires separately deriving both the conditional likelihood and the prior density of the state parameters. However, when the state and measurement equations are correlated, the derivation of the conditional posterior of the state parameters becomes more cumbersome \citep{grant2017bayesian,grant2017reconciling,leiva2023endogenous}.

In this paper, we introduce a novel and direct approach to derive the conditional posterior of $\boldsymbol{\tau}$. The key idea is to first obtain the joint conditional distribution of $(\boldsymbol{\tau}', \mathbf{y}')'$, which will be shown later to be a Gaussian distribution, and then exploit the standard Gaussian conditioning properties to obtain the conditional posterior of $\boldsymbol{\tau}$. Specifically, by stacking equation~\eqref{eq:aug_SVAR} over $t=1,\ldots,T$, we obtain
\begin{align}
	\mathbf{H}_1 \widetilde{\boldsymbol{\eta}} = \boldsymbol{\Xi} \bar{\boldsymbol{\tau}}_0 + \widetilde{\mathbf{u}}, \quad \widetilde{\mathbf{u}} \sim \mathcal{N}(\mathbf{0}, \mathbf{I}_T \otimes \widetilde{\boldsymbol{\Sigma}}), \label{eq:etatilde}
\end{align} where $\widetilde{\boldsymbol{\eta}} = (\widetilde{\boldsymbol{\eta}}_1', \ldots, \widetilde{\boldsymbol{\eta}}_T')'$,  $\widetilde{\boldsymbol{\Sigma}} = \widetilde{\mathbf{B}} \widetilde{\mathbf{B}}$,
\begin{align*}
	\mathbf{H}_1 = 
	\begin{pmatrix}
		\mathbf{I}_7 & \mathbf{0}_7 & \mathbf{0}_7 & \cdots & \cdots & \cdots & \cdots & \mathbf{0}_7 \\
		-\widetilde{\mathbf{A}}_1 & \mathbf{I}_7 & \mathbf{0}_7 & \ddots & \ddots & \ddots & \ddots & \vdots \\
		-\widetilde{\mathbf{A}}_{2} &  -\widetilde{\mathbf{A}}_1 & \mathbf{I}_7 & \mathbf{0}_7 & \ddots & \ddots & \ddots & \vdots \\
		\vdots & \ddots & \ddots & \ddots & \ddots & \ddots & \ddots & \vdots \\
		-\widetilde{\mathbf{A}}_{p} &  \ddots & -\widetilde{\mathbf{A}}_2 & -\widetilde{\mathbf{A}}_1 & \mathbf{I}_7 & \mathbf{0}_7 & \ddots & \vdots \\
		\mathbf{0}_7 &  \ddots & \ddots & \ddots & \ddots  & \ddots & \ddots & \vdots \\
		\vdots  &  \ddots  & \ddots & \ddots & \ddots & \ddots & \ddots & \mathbf{0}_7 \\
		\mathbf{0}_7 & \cdots & \mathbf{0}_7  & -\widetilde{\mathbf{A}}_{p} & \cdots & -\widetilde{\mathbf{A}}_{2} & -\widetilde{\mathbf{A}}_1 &  \mathbf{I}_7
	\end{pmatrix},
	\quad 
	\boldsymbol{\Xi} = 
	\begin{pmatrix}
		\mathbf{I}_3 \\
		\mathbf{0}_{4 \times 3}\\
		\mathbf{0}_{7\times 3} \\
		\vdots \\
		\vdots \\
		\mathbf{0}_{7 \times 3}
	\end{pmatrix}
\end{align*}
Next, given the independent Gaussian prior for $\boldsymbol{\tau}_{0}$ in~\eqref{eq:prior}, we can represent $(\boldsymbol{\tau}_{0}', \widetilde{\boldsymbol{\eta}}')'$ through the following linear system:
\begin{align}
	\mathbf{H}_2
	\begin{pmatrix}
		\bar{\boldsymbol{\tau}}_0 \\ \widetilde{\boldsymbol{\eta}}
	\end{pmatrix}
	= \widetilde{\boldsymbol{\tau}} + \mathbf{e}, \quad \mathbf{e} \sim \mathcal{N} \left(\mathbf{0}, \boldsymbol{\Omega} \right), \label{eq:tau0_etatilde} 
\end{align}
where $\widetilde{\boldsymbol{\tau}} = (\bar{\boldsymbol{\tau}}_{00}', 0, \cdots, 0)'$, $\boldsymbol{\Omega} = \text{diag} (\mathbf{V}_{\bar{\tau}_{00}}, \mathbf{I}_T \otimes \widetilde{\boldsymbol{\Sigma}} )$ and
\begin{align*}
	\mathbf{H}_2 = 
	\begin{pmatrix}
		\mathbf{I}_4 & \mathbf{0}_{4 \times 7T} \\
		-\boldsymbol{\Xi} & \mathbf{H}_1
	\end{pmatrix}.
\end{align*}
Based on the result in \eqref{eq:tau0_etatilde}, we now derive the joint conditional distribution of $(\boldsymbol{\tau}', \mathbf{y}')'$. First, by definition, it can be verified that $\widetilde{\boldsymbol{\eta}}_t$ relates linearly to $(\boldsymbol{\tau}_t', \mathbf{y}_t')'$ as:
\begin{align}
	\widetilde{\boldsymbol{\eta}}_t = \mathbf{Q} \begin{pmatrix} \boldsymbol{\tau}_t \\ \mathbf{y}_t  \end{pmatrix},  \label{eq:etatilde_tauy}
\end{align}
for $t = 1, \ldots, T$, where 
\begin{align*}
	\mathbf{Q} = 
	\begin{pmatrix}
		\mathbf{I}_3 & \mathbf{0}_{3} & \mathbf{0}_{3 \times 1} \\
		\widetilde{\mathbf{Q}} & \mathbf{I}_{3} & \mathbf{0}_{3 \times 1} \\
		\mathbf{0}_{1 \times 3} & \mathbf{0}_{1 \times 3} & 1
	\end{pmatrix},
	\quad
	\widetilde{\mathbf{Q}} =  
	\begin{pmatrix}
		-1 & 0 & 0\\
		0 & -1 & 0 \\
		0 & -1 & -1
	\end{pmatrix}.
\end{align*}
Substituting the linear relationship in \eqref{eq:etatilde_tauy} into \eqref{eq:tau0_etatilde} yields
\begin{align*}
	\mathbf{H} \mathbf{z}
	=
	\widetilde{\boldsymbol{\tau}} + \mathbf{e}, \quad \mathbf{e} \sim \mathcal{N} \left(\mathbf{0}, \boldsymbol{\Omega} \right),
\end{align*}
where $\mathbf{z} = (\bar{\boldsymbol{\tau}}_0', \boldsymbol{\tau}_1', \mathbf{y}_1', \ldots, \boldsymbol{\tau}_T', \mathbf{y}_T')'$ and
\begin{align*}
	\mathbf{H} = \mathbf{H}_{2} 
	\begin{pmatrix}
		\mathbf{I}_4 & \mathbf{0}_{4 \times 7T} \\
		\mathbf{0}_{7T \times 4} & \mathbf{I}_T \otimes \mathbf{Q}
	\end{pmatrix}.
\end{align*} 
This suggests that $\mathbf{z} \sim \mathcal{N}( \boldsymbol{\mu}_z, \mathbf{K}_z^{-1} )$ with $\boldsymbol{\mu}_z = \mathbf{H}^{-1} \widetilde{\boldsymbol{\tau}}$ and $\mathbf{K}_z = \mathbf{H}' \boldsymbol{\Omega}^{-1} \mathbf{H}$. 

\begin{figure}[H]
	\centering
	\includegraphics[width=0.5\linewidth]{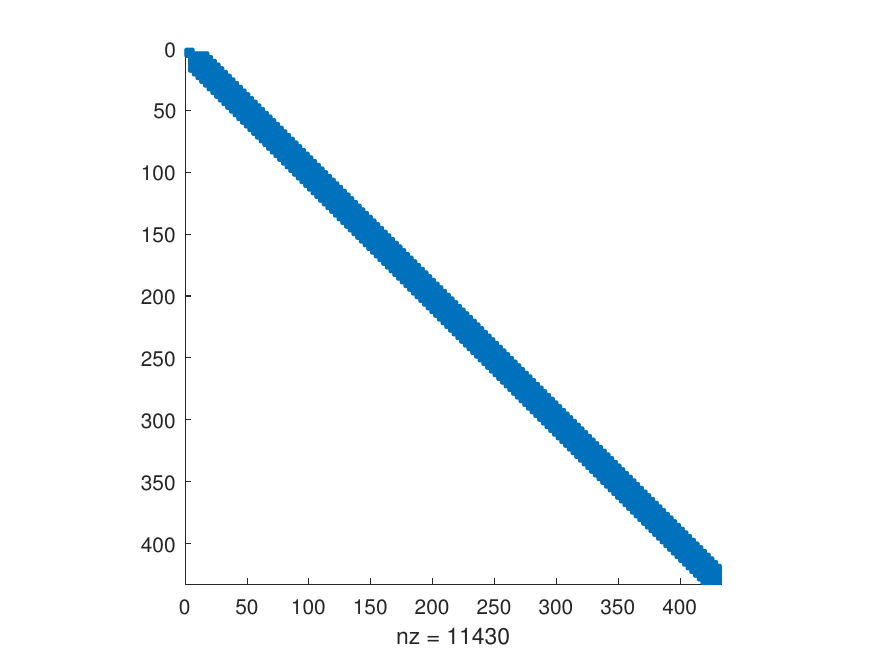}
	\caption{The sparsity pattern of the precision matrix $\mathbf{K}_{\tau}$.}
	\label{fig:sparseKtau}
\end{figure}

Note that since $\mathbf{z}$ has a Gaussian distribution and contains the same elements as $(\boldsymbol{\tau}', \mathbf{y}')'$ up to permutation, the standard properties of Gaussian distribution imply that $(\boldsymbol{\tau}', \mathbf{y}')'$ is also Gaussian. More precisely, let $\mathbf{P}_z$ be a permutation matrix such that $\mathbf{P}_z \mathbf{z} = (\boldsymbol{\tau}', \mathbf{y}')'$. Then we have the following result:
\begin{align}
	\left( (\boldsymbol{\tau}', \mathbf{y}')'  | \mathbf{B}, \beta, \alpha, \boldsymbol{\Phi}, \kappa_{1}, \kappa_{2} \right) \sim \mathcal{N}( \boldsymbol{\mu} ,  \mathbf{K}^{-1} ), \label{eq:joint_tau_y}
\end{align}
where $ \boldsymbol{\mu} = \mathbf{P}_z \boldsymbol{\mu}_z$ and $\mathbf{K} = \mathbf{P}_z \mathbf{K}_z \mathbf{P}_z'$. Finally, we partition the mean vector $\boldsymbol{\mu}$ and precision matrix $\mathbf{K}$ according to the dimensions of $\boldsymbol{\tau}$ and $\mathbf{y}$ as
\begin{align*}
	\boldsymbol{\mu} = 
	\begin{pmatrix}
		\boldsymbol{\mu}_{\tau} \\ \boldsymbol{\mu}_y
	\end{pmatrix},
	\quad 
	\mathbf{K} = 
	\begin{pmatrix}
		\mathbf{K}_{\tau} & \mathbf{K}_{\tau,y} \\
		\mathbf{K}_{\tau,y}' & \mathbf{K}_{y}
	\end{pmatrix}.
\end{align*}
By applying the standard Gaussian conditioning results, the conditional posterior of $\boldsymbol{\tau}$ is given by
\begin{align*}
	(\boldsymbol{\tau} | \mathbf{B}, \beta, \alpha, \boldsymbol{\Phi}, \kappa_{1}, \kappa_{2},\mathbf{y}) \sim \mathcal{N}( \widehat{\boldsymbol{\tau}}, \mathbf{K}_{\tau}^{-1}),
\end{align*}
where $\widehat{\boldsymbol{\tau}} = \boldsymbol{\mu}_{\tau} - \mathbf{K}_{\tau}^{-1} \mathbf{K}_{\tau, y} ( \mathbf{y} - \boldsymbol{\mu}_y )$. Note that the $(3T+3) \times (3T+3)$ precision matrix $\mathbf{K}_{\tau}$  is sparse and banded, as illustrated in Figure~\ref{fig:sparseKtau}, which visualizes its structure with nonzero and zero elements shown in blue and white, respectively.\footnote{Here the dimension of $\mathbf{K}_{\tau}$ is $432 \times 432$, which matches the dimension used in our empirical application in Section~\ref{sec:Empirical Analysis}.} This banded structure enables efficient sampling from $\mathcal{N}(\widehat{\boldsymbol{\tau}}, \mathbf{K}_{\tau}^{-1})$ using the precision-based approach of \cite{chan2009efficient}. The procedure is as follows: first, we compute the lower triangular Cholesky factor $\mathbf{L}$ of $\mathbf{K}_{\tau}$ such that $\mathbf{K}_{\tau} = \mathbf{L} \mathbf{L}'$, which is computationally efficient due to the banded structure of $\mathbf{K}_{\tau}$. A draw $\boldsymbol{\tau} \sim \mathcal{N}(\widehat{\boldsymbol{\tau}}, \mathbf{K}_{\tau}^{-1})$ is then obtained by solving:
\begin{align*}
	\boldsymbol{\tau}  = \boldsymbol{\mu}_{\tau} - \mathbf{L}^{' -1} \left(  \mathbf{L}^{-1} \mathbf{K}_{\tau, y} ( \mathbf{y} - \boldsymbol{\mu}_y ) + \mathbf{f}  \right), \quad \mathbf{f} \sim \mathcal{N}\left( \mathbf{0}_{3(T+1) \times 1}, \mathbf{I}_{3(T+1)} \right).
\end{align*}
Since $\mathbf{L}$ is a banded lower triangular matrix, we can solve the required linear systems rapidly via forward and backward substitution, with $\mathcal{O}( T )$ complexity. This efficiently avoids directly computing $\mathbf{K}_{\tau}^{-1}$, which has $\mathcal{O}( T^3 )$ complexity.

For sampling $(\mathbf{B}, \beta, \alpha)$ in Step 2, we first note that these parameters correspond to the nonzero elements in the contemporaneous impact matrix $\widetilde{\mathbf{B}}$ of the augmented structural autoregression representation in \eqref{eq:aug_SVAR}. Consequently, sampling $(\mathbf{B}, \beta, \alpha)$ is equivalent to sampling $\widetilde{\mathbf{B}}$ with the zero restrictions in \eqref{eq:zero_B} and the sign restriction on $\alpha$. Since these restrictions can be expressed as linear equality and inequality restrictions on $\widetilde{\mathbf{B}}$, we sample the nonzero parameters using the efficient sampler developed by \cite{hou2024large}. We refer readers to this paper for more implementation details.

Since Step 3 - Step 5 of the posterior sampler implement either standard Bayesian procedures or minor modifications of established methods, we relegate their technical details to Appendix~A.

\section{Marginal Likelihood Estimation} \label{sec: ML}

Marginal likelihood is the standard criterion for Bayesian model comparison. In this section, we present an approach for estimating the marginal likelihood of our proposed SMUC-IV  and its restricted versions. Our method builds upon the modified harmonic mean estimator proposed by \cite{gelfand1994bayesian}, integrating it with the conditional Monte Carlo method of \cite{chan2023comparing} to enhance estimation accuracy. We begin with an overview of the modified harmonic mean estimator and then outline our conditional Monte Carlo improved estimator for the SMUC-IV. Technical details are relegated to Appendix~B.

\subsection{Modified Harmonic Mean Estimator}
The marginal likelihood of a given model is defined as 
\begin{align*}
	p(\mathbf{y}) = \int p( \mathbf{y} | \boldsymbol{\theta}) p( \boldsymbol{\theta}) d \boldsymbol{\theta},
\end{align*} 
where $\mathbf{y}$ denotes a vector of observed data, $\boldsymbol{\theta}$ denotes the set of all parameters specific to the model, $p( \mathbf{y} | \boldsymbol{\theta})$ is the likelihood function, and $p(\boldsymbol{\theta})$ is the prior density for the model. The modified harmonic mean estimator is built upon the following identity:
\begin{align*}
	p(\mathbf{y})^{-1} = \int \frac{q(\boldsymbol{\theta})}{p( \mathbf{y} | \boldsymbol{\theta}) p(\boldsymbol{\theta})} p(\boldsymbol{\theta} | \mathbf{y}) d \boldsymbol{\theta}.
\end{align*}
Here $p(\boldsymbol{\theta} | \mathbf{y})$ is the posterior distribution and $q(\boldsymbol{\theta})$ is a tuning function that can be any density function defined on $\boldsymbol{\theta}$ with its support contained in the support of the posterior density, i.e., $q(\boldsymbol{\theta}) > 0$ implies $p(\boldsymbol{\theta}|\mathbf{y}) > 0$. This suggests that the marginal likelihood can be estimated using the following estimator:
\begin{align*}
	\widehat{p(\mathbf{y})}_{GD} = \left( \frac{1}{R} \sum_{i = 1}^{R}  \frac{q(\boldsymbol{\theta}^{(i)}  )}{p( \mathbf{y} | \boldsymbol{\theta}^{(i)}) p(\boldsymbol{\theta}^{(i)})}  \right)^{-1},
\end{align*}
where $\boldsymbol{\theta}^{(1)}, \ldots, \boldsymbol{\theta}^{(R)}$ are draws from the posterior distribution $p(\boldsymbol{\theta} | \mathbf{y})$. 

While the modified harmonic mean estimator described above is simulation-consistent and straightforward to implement, it might perform poorly when the model parameter $\boldsymbol{\theta}$ is high-dimensional. To improve estimation accuracy, we employ the conditional Monte Carlo method of \cite{chan2023comparing}. The key idea is to first analytically integrate out as many parameters in $\boldsymbol{\theta}$ as possible, and then construct a modified harmonic mean estimator using the resulting unconditional likelihood and prior on the remaining lower-dimensional parameters.

\subsection{Estimating the Marginal Likelihood for SMUC-IV}
We now outline our conditional Monte Carlo improved modified harmonic mean estimator for the marginal likelihood of the SMUC-IV.  In our modelling setting, the marginal likelihood is given by
\begin{align*}
	p(\mathbf{y})&= \int p(\mathbf{y} | \boldsymbol{\tau}, \boldsymbol{\Phi}, \mathbf{B}, \beta, \alpha, \kappa_1, \kappa_2) p(\boldsymbol{\tau}, \boldsymbol{\Phi}, \mathbf{B}, \beta, \alpha, \kappa_1, \kappa_2) d(\boldsymbol{\tau}, \boldsymbol{\Phi}, \mathbf{B}, \beta, \alpha, \kappa_1, \kappa_2)\\
	&= \int p(\mathbf{y} | \boldsymbol{\tau}, \boldsymbol{\Phi}, \mathbf{B}, \alpha, \beta)p(\boldsymbol{\tau}|\boldsymbol{\Phi}, \mathbf{B}, \alpha, \beta) p(\boldsymbol{\Phi} | \kappa_1, \kappa_2) p(\kappa_1) p(\kappa_2) p(\mathbf{B}) p(\alpha) p (\beta) d(\boldsymbol{\tau}, \boldsymbol{\Phi}, \mathbf{B}, \beta, \alpha, \kappa_1, \kappa_2)\\
	&= \int p( \mathbf{y} | \boldsymbol{\Phi}, \mathbf{B}, \alpha, \beta) p(\mathbf{B}) p(\boldsymbol{\Phi}) p(\alpha)p(\beta) d(\boldsymbol{\Phi}, \mathbf{B}, \beta, \alpha).
\end{align*}
The second equality follows from the conditional independence structure of the prior distributions. In the last equality, we have integrated out the state parameters $\boldsymbol{\tau}$ and the hyperparameters $(\kappa_1, \kappa_2)$. Specifically, we have:
\begin{align}
	&p( \mathbf{y} | \boldsymbol{\Phi}, \mathbf{B}, \alpha, \beta) = \int p(\mathbf{y} | \boldsymbol{\tau}, \boldsymbol{\Phi}, \mathbf{B}, \alpha, \beta)p(\boldsymbol{\tau}|\boldsymbol{\Phi}, \mathbf{B}, \alpha, \beta) d \boldsymbol{\tau}, \label{eq:ML_tau} \\
	&p(\boldsymbol{\Phi}) = \int p(\boldsymbol{\Phi} | \kappa_1, \kappa_2) p(\kappa_1) p(\kappa_2) d(\kappa_1, \kappa_2). \label{eq:ML_kappa}
\end{align}
Note that the expression in \eqref{eq:ML_tau} can be obtained directly from the result in \eqref{eq:joint_tau_y} by using the property that the marginal distribution of joint Gaussian variates is Gaussian. To be specific, let $\boldsymbol{\Lambda} = \mathbf{K}^{-1}$ be the covariance matrix of the joint conditional distribution of $(\boldsymbol{\tau}', \mathbf{y}')'$, and partition $\boldsymbol{\Lambda}$ according to the dimensions of $\boldsymbol{\tau}$ and $\mathbf{y}$ as
\begin{align*}
	\boldsymbol{\Lambda} = 
	\begin{pmatrix}
		\boldsymbol{\Lambda}_{\tau} & \boldsymbol{\Lambda}_{\tau,y} \\
		\boldsymbol{\Lambda}_{\tau,y}' & \boldsymbol{\Lambda}_y
	\end{pmatrix}.
\end{align*} 
A standard property of the Gaussian distribution implies that the marginal distribution of $\mathbf{y}$ (of dimension $4T \times 1$) is a Gaussian distribution with mean vector $\boldsymbol{\mu}_y$ and covariance matrix $\boldsymbol{\Lambda}_y$. This yields
\begin{align*}
	p( \mathbf{y} | \boldsymbol{\Phi}, \mathbf{B}, \alpha, \beta) = (2 \pi)^{- 2T} | \boldsymbol{\Lambda}_y  |^{-\frac{1}{2}} e^{-\frac{1}{2} \left(  \mathbf{y} -  \boldsymbol{\mu}_y   \right)'  \boldsymbol{\Lambda}^{-1}_y  \left( \mathbf{y} -  \boldsymbol{\mu}_y  \right)  }.
\end{align*}
For the analytical expression of the marginal prior $p(\boldsymbol{\Phi})$ in \eqref{eq:ML_kappa}, we refer readers to Appendix~B for details.

Given the analytical expressions for \eqref{eq:ML_tau} and \eqref{eq:ML_kappa}, we can estimate the marginal likelihood of our SMUC-IV with the following conditional Monte Carlo improved modified harmonic mean estimator:
\begin{align}
	\widehat{p(\mathbf{y})}_{CMGD} = \left( \frac{1}{R} \sum_{i=1}^{R} \frac{ q(\boldsymbol{\Phi}^{(i)}, \mathbf{B}^{(i)}, \alpha^{(i)}, \beta^{(i)}) }{p( \mathbf{y} | \boldsymbol{\Phi}^{(i)}, \mathbf{B}^{(i)}, \alpha^{(i)}, \beta^{(i)}) p(\mathbf{B}^{(i)}) p(\boldsymbol{\Phi}^{(i)}) p(\alpha^{(i)})p(\beta^{(i)})} \right)^{-1}, \label{eq:CMGD}
\end{align}
where $(\boldsymbol{\Phi}^{(i)}, \mathbf{B}^{(i)}, \alpha^{(i)}, \beta^{(i)})$, $i = 1, \ldots, R$, are posterior draws that can be obtained using the posterior sampler described in the last section. Note that the key difference between this estimator and the standard modified harmonic mean estimator is the use of the likelihood function $p(\mathbf{y}|\boldsymbol{\Phi}, \mathbf{B}, \alpha, \beta)$, which is unconditional on the high-dimensional state parameters $\boldsymbol{\tau}$, and the marginal prior $p(\boldsymbol{\Phi})$. This formulation substantially reduces the dimensionality of the numerical integration, which in turn reduces the Monte Carlo variance of the estimator, resulting in greater numerical stability and estimation precision (see \cite{chan2023comparing} for more discussion).

We now turn to the choice of the tuning density function $q(\boldsymbol{\Phi}, \mathbf{B}, \alpha, \beta)$, a critical component of our estimator \eqref{eq:CMGD} that is essential for the accuracy of marginal likelihood estimation. Although the posterior density is the theoretically optimal choice, it is computationally intractable. Therefore, we follow \cite{chan2023comparing} and approximate the posterior using a truncated Gaussian density. Specifically, we consider a tuning density function that takes the following form:
\begin{align}
	q( \boldsymbol{\Phi}, \mathbf{B}, \alpha, \beta) = q_{\Phi}(\boldsymbol{\Phi}) q_{B}(\mathbf{B}) q_{\alpha}(\alpha) q_{\beta}(\beta),
\end{align}
where each $q_{j}(\cdot)$ for $j \in \{\Phi, B, \alpha, \beta\}$ is a truncated Gaussian density with mean and covariance matrix set to the estimated posterior mean and covariance matrix.\footnote{ Using an appropriately truncated tuning density can ensure that the modified harmonic mean estimator has a finite variance \citep{geweke1999using}. } The full details for $q_{\Phi}(\boldsymbol{\Phi})$, $q_{B}(\mathbf{B})$, $q_{\alpha}(\alpha)$, and $q_{\beta}(\beta)$ are provided in Appendix~B.

\section{Empirical Analysis}\label{sec:Empirical Analysis}
In this section, we first describe the data in Section~\ref{Sec:data} and then conduct a model comparison exercise in Section~\ref{sec:Model Comparison Exercise} to validate our proposed SMUC-IV framework. In particular, this exercise provides strong evidence in favor of the influence of monetary policy shocks on the stars. Section~\ref{sec:Estimates of Macroeconomic Stars} presents our estimates of the macroeconomic stars---namely, the potential GDP growth, trend inflation, and the neutral interest rate. Section~\ref{sec:effects of MP} examines the effects of monetary policy shocks on these stars. Subsequently, Section~\ref{sec:HD} assesses the role of the monetary policy  in driving the evolution of the macroeconomic stars. Section \ref{sec:results discussion} provides a discussion of the results and their policy implications.  Finally, robustness analyses are reported in Section~\ref{sec:Robustness checks}.


\subsection{Data and External Instrument}\label{Sec:data}
We estimate the empirical model using the following quarterly data from 1988:Q1 to 2023:Q3: real GDP, the GDP deflator, the federal funds effective rate, the shadow interest rate, and the orthogonal monetary policy surprise \citep{bauer2023reassessment}. To capture both conventional and unconventional monetary policy
when the nominal federal funds rate is at the effective lower bound (ELB), we use the \cite{wu2016measuring} shadow short rate. Consistent with the quarterly frequency of our data, we use four lags in our model. We transform real GDP as $100 \log(x_t)$ and compute real GDP growth as the first difference of this series. The quarterly GDP deflator inflation rate is measured as annualized log growth, 400 $\log(x_t / x_{t-1})$. The federal funds effective rate, shadow interest rate, and orthogonal monetary policy surprise are originally monthly series; we obtain their quarterly versions by averaging the three monthly observations within each quarter. Real GDP, the GDP deflator, and the federal funds effective rate are available from the FRED database. We consider different measures of inflation and the interest rate as well as consider the bond premium \citep{favara2016updating} as a control variable in the vector autoregression, see section \ref{sec:Robustness checks}. The shadow interest rate is from the website of the Federal Reserve Bank of Atlanta, the excess bond premium is from the website of the Board of Governors of the Federal Reserve System, and the orthogonal monetary policy surprise series is from the website of the Federal Reserve Bank of San Francisco. 

In this paper, we use the orthogonal monetary policy surprise series developed by \citet{bauer2023reassessment} as our external instrument. This series identifies monetary policy shocks from high-frequency asset price movements in narrow windows around policy announcements. High-frequency interest rate changes around FOMC announcements are a common tool for identifying monetary policy effects, but recent studies have questioned their exogeneity and relevance as instruments, particularly for estimating macroeconomic impacts \citep[e.g.,][]{ramey2016macroeconomic,miranda2021transmission}. For instance, monetary policy surprises may be correlated with publicly available macroeconomic and financial data released before FOMC announcements. To address these concerns, \citet{bauer2023reassessment} (i) expand the set of monetary policy events to include speeches by the Federal Reserve Chair---roughly doubling the number of announcements---and (ii) orthogonalize the resulting  surprises with respect to pre-announcement macroeconomic and financial data to account for predictability via the ``Fed response to news" channel.
In Section~\ref{sec:Robustness checks}, we also conduct a robustness analysis by considering an alternative instrument in our application, and we find that our main empirical results remain robust under this alternative choice of instrument.


\subsection{Model Comparison Exercise}\label{sec:Model Comparison Exercise}
This section conducts a Bayesian model comparison exercise by evaluating the marginal likelihoods of alternative specifications of our SMUC-IV.

The first alternative specification imposes the restriction that monetary policy shocks have no contemporaneous effects on all of the trend components, that is, the macroeconomic stars of interest in this paper. We refer to this model as SMUC-IV-R1. Next, we consider a restricted version of our SMUC-IV model that imposes zero correlation between the external instrument and all structural shocks by setting $\beta = 0$; we refer to this specification as SMUC-IV-R2.  A comparison between our proposed SMUC-IV and SMUC-IV-R2 serves to test the relevance condition of the instrument. Note that under the restriction $\beta = 0$, the proxy equation \eqref{eq:proxy} is independent of the system of equations in the unobserved components model specified in \eqref{eq:SVAR}. Therefore, by imposing various patterns of zero restrictions on $\mathbf{B}$, we can assess different correlation structures between the trend and cycle innovations. In our model comparison exercise, we further consider two nested versions of SMUC-IV-R2. The first nested version of SMUC-IV-R2 is denoted as SMUC-IV-R3. This specification assumes that innovations between the trend and cycle components are uncorrelated, while allowing for correlation within the innovations of the trend components and within the innovations of the cycle components, respectively. Specifically, under SMUC-IV-R3 we assume the $6 \times 6$ contemporaneous response matrix $\mathbf{B}$ to be a block diagonal matrix with each block of dimension $3 \times 3$. For the second nested version of SMUC-IV-R2, we assume $\mathbf{B}$ to be a diagonal matrix and denote this model as SMUC-IV-R4. This specification is similar to that in many conventional studies on multivariate unobserved components models, which assumes all innovations of the trend and cycle components are mutually independent. Table~\ref{table:competing_model} provides a summary of the competing models in our model comparison exercise.

\begin{table}[H]
	\caption{Competing models used in the comparison exercise.}
	\label{table:competing_model}
	\centering
	\footnotesize
	\begin{tabular}{l p{0.68\linewidth}}
		\hline\hline
		Model & Description \\
		\hline
		SMUC-IV    & The baseline model specified in equations \eqref{eq:aug_SVAR} - \eqref{eq:zero_B}. \\
		SMUC-IV-R1 & A restricted version of SMUC-IV by imposing zero contemporaneous effects of monetary policy shocks on the trend components. \\
		SMUC-IV-R2    &  A restricted version of the SMUC-IV that imposes zero correlation between the external instrument and all structural shocks by setting $\beta=0$. \\
		SMUC-IV-R3    & A restricted version of the SMUC-IV-R2 that imposes zero correlation between trend and cycle innovations \\
		SMUC-IV-R4      & A restricted version of the SMUC-IV-R2 that imposes zero correlation between all trend and cycle innovations. \\
	
		\hline\hline
	\end{tabular}
\end{table}

Table~\ref{tab:log-ML-combined} reports the log marginal likelihood estimates. The results provide significant evidence that our proposed SMUC-IV model outperforms all alternative specifications considered. A few findings are also worth highlighting. First, by comparing the SMUC-IV with the SMUC-IV-R1, the difference in the log marginal likelihoods is about $18$, strongly supporting the non-zero contemporaneous effects of monetary policy shocks on the trend components. Second, the SMUC-IV-R2 performs better than the SMUC-IV-R3 and SMUC-IV-R4. This suggests that allowing for correlated innovations among all trend and cycle components is an important modelling feature supported by the data. In particular, the log marginal likelihood of SMUC-IV-R2 is about $22$ higher than that of SMUC-IV-R4, indicating that the assumption of independent innovations for the trend and cycle components is empirically implausible in our empirical analysis. Lastly, the SMUC-IV outperforms SMUC-IV-R2, suggesting that the relevance condition of the instrument is satisfied.

\begin{table}[H]
	\caption{Estimated log marginal likelihoods}
	\label{tab:log-ML-combined}
	\centering
	\footnotesize
	\begin{tabular}{ccccc}
		\hline\hline
		SMUC-IV & SMUC-IV-R1 & SMUC-IV-R2& SMUC-IV-R3& SMUC-IV-R4 \\
		\hline
		\textbf{-205} & -223 & -208 & -225 & -230 \\
		\hline\hline
	\end{tabular}
\end{table}

\subsection{Estimates of Macroeconomic Stars}\label{sec:Estimates of Macroeconomic Stars}
Although macroeconomic stars---for example, the neutral interest rate---have been defined in various ways across the literature, most theoretical frameworks suggest that shifts in the low-frequency components of macroeconomic variables are closely linked to movements in any theoretically defined star.  Accordingly, we examine the persistent evolution of macroeconomic stars through the lens of trend dynamics.  Using US data, we jointly estimate three stars: the  potential GDP growth, $g^{*}$, the trend inflation, $\pi^{*}$, and the  neutral  interest rate, $r^{*}$.   

\begin{figure}[H]
	\centering
	\includegraphics[width=1\linewidth]{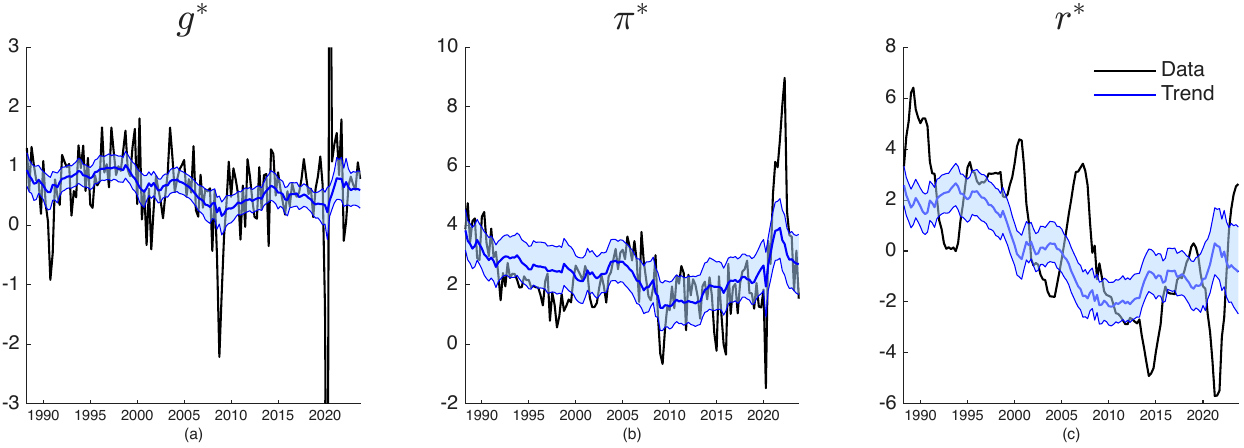}
	\caption{The plot compares the actual data with trend estimates along 68\% credible bands.}
	\label{fig:trends}
\end{figure}

Figure \ref{fig:trends} displays the posterior means of the three estimated stars---$g^{*}$, $\pi^{*}$, $r^{*}$. The trend components provide smooth and plausible estimates of macroeconomic stars. Our results align with the broader literature, capturing the common tendencies documented across studies.

Panel (a) of Figure \ref{fig:trends} shows estimates of $ g^{*}$. From the early to late 1990s, $ g^{*}$ rises noticeably, likely reflecting the internet technology boom. After the 1990s, however, it follows a downward trend until around 2010. Then it follows an upward trend until the early 2020s. Our estimates show a similar tendency to those in \citep{grant2017bayesian,grant2017reconciling,maffei2025identifying}. Sharp declines are evident during major crises such as the dot-com bust in the early 2000s, the Global Financial Crisis of 2007--2009, and the COVID-19 pandemic. These episodes coincide with well-documented hysteresis effects, where severe recessions leave lasting scars on the economy's productive capacity \citep{cerra2005did,cerra2008growth,cerra2023hysteresis}.

Panel (b) of Figure \ref{fig:trends} reports the estimates of $\pi^{*}$. Consistent with prior work \citep[e.g.,][]{stockwatson2,stock2016core,chan2018new,eo2023understanding}, $\pi^{*}$ displays remarkable stability: after falling in the 1990s, it remains anchored near 2\% from the 2000s until the pandemic period. This pattern highlights the Federal Reserve's success in stabilizing long-run inflation expectations.

Finally, Panel (c) of Figure \ref{fig:trends} presents estimates of $r^{*}$. The estimates show a steady decline from the 1990s onward, followed by relative stability during the 2010s-2020s. Our estimates of $r^{*}$ capture the secular decline in the neutral interest rate documented in the literature on estimating $r^{*}$ \citep[e.g.,][]{laubach2003measuring,lubik2015calculating,del2017safety,morley2024simple}.

\begin{figure}[H]
	\centering
	\includegraphics[width=1\linewidth]{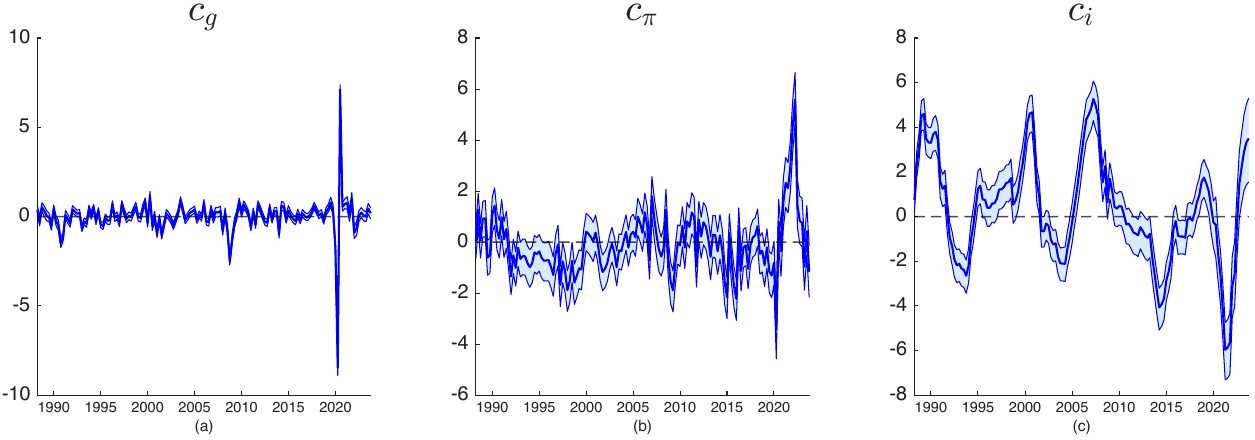}
	\caption{The plot shows cycle estimates with 68\% credible bands.}
	\label{fig:cyclesbaseline}
\end{figure}

We also report the estimated cyclical components in Figure \ref{fig:cyclesbaseline}. Overall, the cycles appear stationary, suggesting that our proposed model effectively decomposes the time series of interest into trends and cycles.  The cyclical components
capture major economic fluctuations, including the recessions in the early 1990s, the global financial crisis of 2008-2009, and the sharp contraction during the COVID-19 pandemic in 2020. 

\subsection{The Effects of Monetary Policy Shocks on the Macroeconomic Stars}\label{sec:effects of MP}

Having obtained plausible estimates of the macroeconomic stars, we now examine the effects of monetary policy shocks on $g^{*}$, $\pi^{*}$, and $r^{*}$. We identify monetary policy shocks using an external instrument based on the high-frequency policy surprises constructed by \cite{bauer2023reassessment}.

\begin{figure}[H]
	\centering
	\includegraphics[width=1\linewidth]{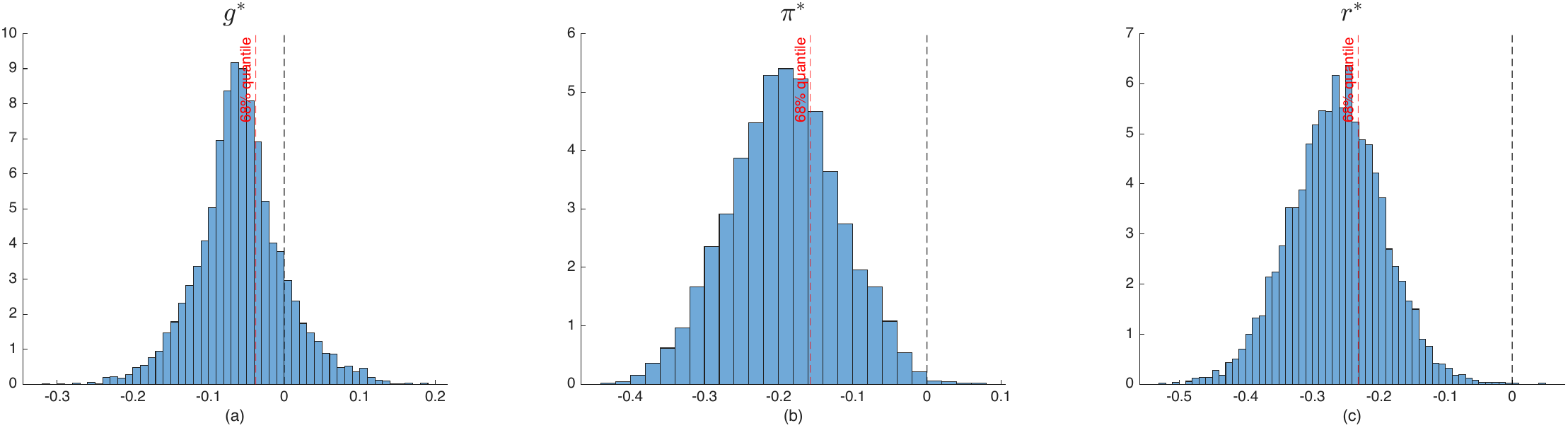}
	\caption{The plot shows the posterior distributions of the impulse response of the trends to a monetary policy shock. Note that the responses are constant over time. }
	\label{fig:IRFtrends}
\end{figure}

\begin{figure}[H]
	\centering
	\includegraphics[width=1\linewidth]{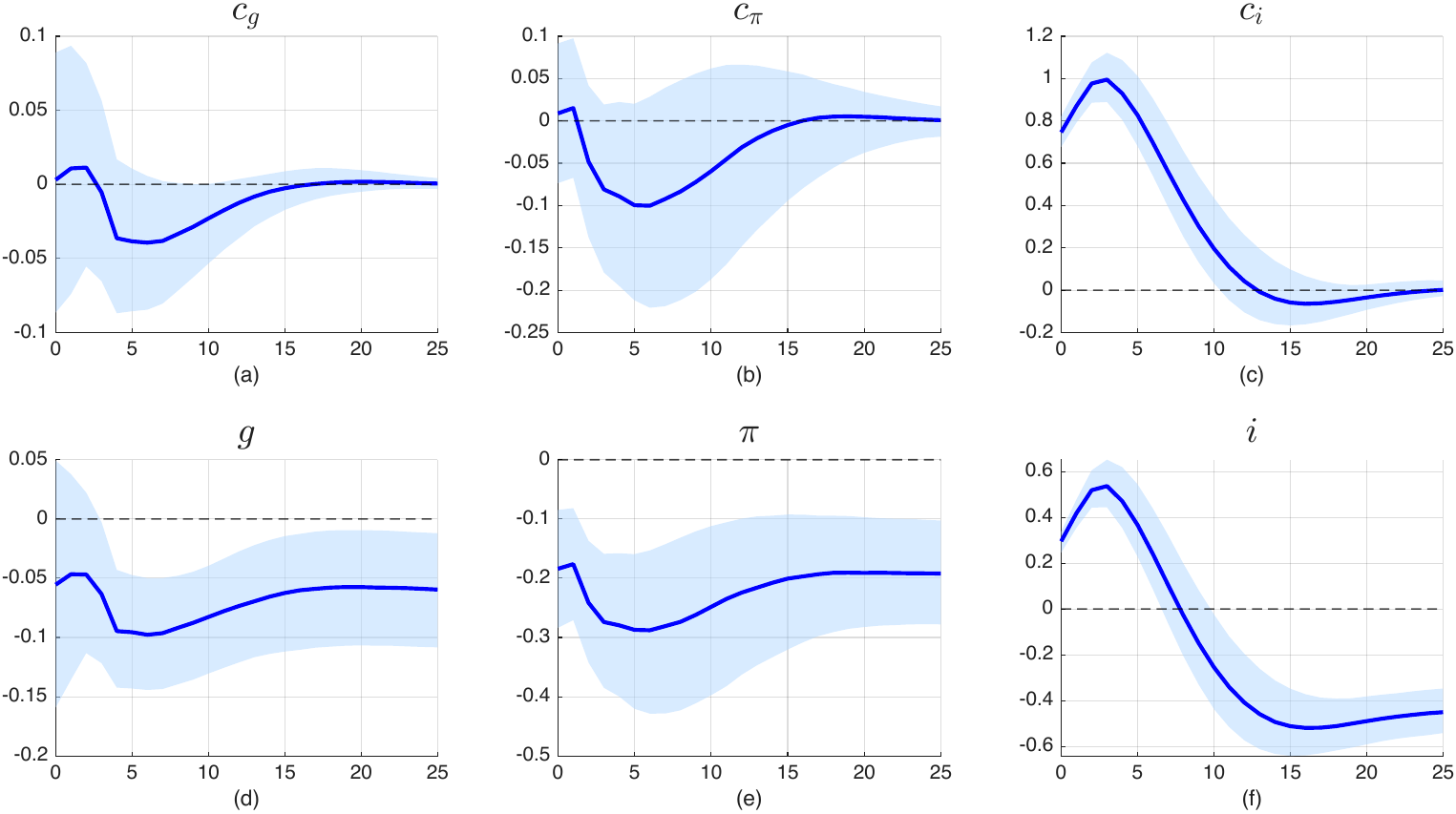}
	\caption{The plots show the impulse responses of the cyclical components and the observed variables to a monetary policy shock, along with 68\% credible intervals.}
	\label{fig:IRFs}
\end{figure}

Figure \ref{fig:IRFtrends} shows the posterior distributions of the impulse responses of $ g^{*}$, $\pi^{*}$, and $r^{*}$ to a one-standard-deviation contractionary monetary policy shock. This shock raises the median response of the nominal interest rate by about 29 basis points on impact (see Figure \ref{fig:IRFs}(f)). By construction, the responses are constant over time, as the model assumes the stars follow a random walk.

Panel (a) of Figure \ref{fig:IRFtrends} shows that $ g^{*}$ falls by about 0.05 percentage points. The negative effects on potential GDP growth align with the empirical evidence in \cite{moran2018innovation}, \cite{garga2021output}, \cite{ma2023monetary}, \cite{jorda2024long}, and \cite{meier2024monetary}, which suggests that tighter monetary policy restrains innovation investment and, in turn, productivity. Panel (b) of Figure \ref{fig:IRFtrends} shows that contractionary monetary policy shocks also reduce $\pi^{*}$ by about 0.2 percentage points. The decline in trend inflation is consistent with evidence that contractionary monetary policy shocks reduce long-term inflation expectations \citep[e.g.,][]{jarocinski2020deconstructing,diegel2021long}. Panel (c) shows that $r^{*} $ falls by about 25 basis points. The decline in the neutral interest rate after contractionary monetary policy shocks could be interpreted as reflecting the  adverse effects of monetary tightening on innovation investment.

As a complement to our analysis of the effects of monetary policy shocks on the stars, we also report the cyclical responses and the responses of the observed variables to contractionary monetary policy shocks in Figure \ref{fig:IRFs}.

Figure \ref{fig:IRFs} shows that the impulse responses align with standard macroeconomic predictions: the nominal interest rate rises, real GDP growth falls, and inflation declines. The figure also indicates that contractionary monetary policy shocks have more persistent effects on the business cycle than expected, taking about 3--5 years to fade out. Moreover, the responses of the observed variables appear to be largely permanent. 

Overall, our results suggest that monetary tightening may generate medium-run adverse effects on aggregate demand, which in turn suppress investment in innovation and weaken long-run productivity, leaving potentially permanent scars on the economy.

\subsection{The Role of Monetary Policy Shocks in Shaping Macroeconomic Stars}\label{sec:HD}

\begin{figure}[H]
	\centering
	\includegraphics[width=1\linewidth]{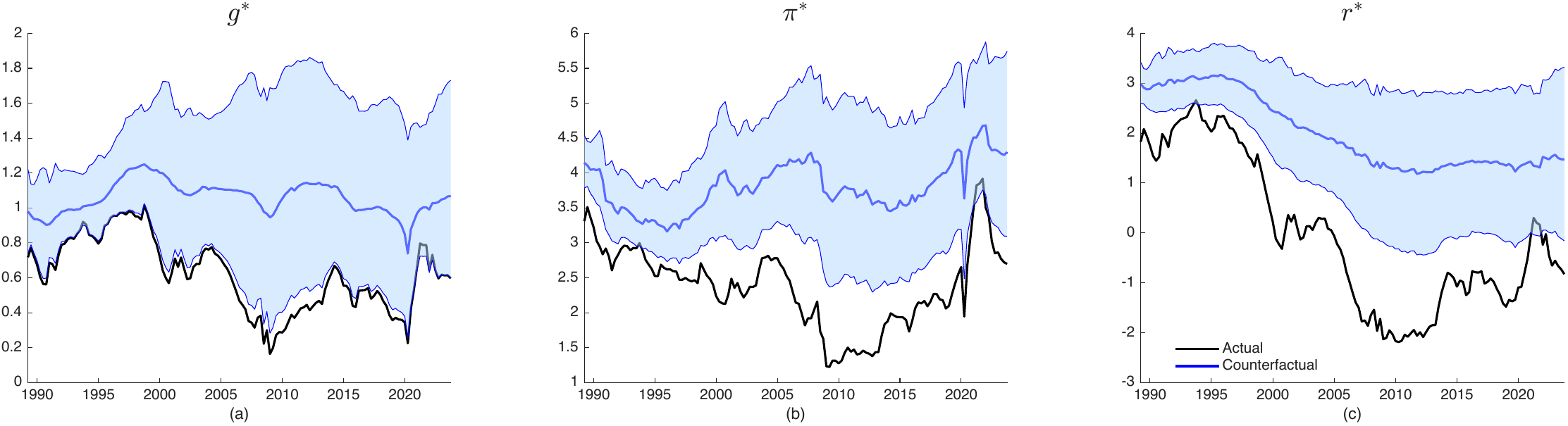}
	\caption{The plot compares the actual trends and the counterfactual path with 68\% credible bands. The counterfactual trends are calculated by shutting down the effects of all monetary policy shocks as is done in the calculation of historical decompositions, see \cite{kilian2017structural}.}
	\label{fig:histdecomp}
\end{figure}

Given the meaningful effects of monetary policy shocks on the macroeconomic stars, an important question is whether these shocks have played an important role in shaping the historical dynamics of the stars. To address this, we conduct a counterfactual analysis based on the historical decomposition of the trend components, as shown in Figure \ref{fig:histdecomp}.

Figure \ref{fig:histdecomp} compares the actual trend paths with counterfactual paths, which are generated by shutting down all identified monetary policy shocks, following the methodology outlined in \cite{kilian2017structural}. The shaded regions represent 68\% credible bands around the counterfactual estimates.

Our counterfactual analysis quantifies the contributions of monetary policy shocks over time. For instance, in Figure \ref{fig:histdecomp}, Panel (a) shows that, absent these shocks, $g^{*}$ would have been over 0.8 percentage points higher during the crisis. Panel (b) shows that $\pi^{*}$, which has remained below its counterfactual path since the early 1990s, would have hovered closer to 3.5 percent in the absence of monetary tightening. Panel (c) shows that $r^{*}$ would have been around 3 percentage points higher during the financial crisis.

%

\subsection{Results Discussion and  Policy Implications}\label{sec:results discussion}

  The negative effects of contractionary monetary policy shocks on $g^{*}$ and $r^{*}$ shown in Figure \ref{fig:IRFtrends} can be rationalized by the mechanism emphasized in the Keynesian endogenous growth model with nominal rigidities in \cite{fornaro2023scars}.  In that setting, long-run growth is endogenous and can be influenced by both demand- and supply-side forces. With nominal rigidities, output can deviate from potential, so monetary policy has real effects.  In their framework, monetary tightening can trigger a ``supply--demand doom loop": a contraction reduces aggregate demand, compressing firms' profits and weakening incentives to invest. The resulting decline in investment lowers expected productivity growth and household wealth, which further depresses demand and induces additional cutbacks in investment. As a result, the $g^{*}$ and $r^{*}$ fall.   Moreover, by amplifying the effects of adverse disturbances on the real economy, this vicious spiral helps interpret panel (a) and (c) in Figure \ref{fig:histdecomp}: while the actual paths of $g^{*}$ and $r^{*}$ (with contractionary monetary policy shocks) and the counterfactual paths (excluding the shocks' effects) share a similar downward tendency, the actual series remain persistently below their counterfactual counterparts.  
 
 Turning to trend inflation, at first glance, significant negative effects of monetary policy shocks in panel (b) of Figure \ref{fig:IRFtrends} may appear undesirable, as they could be read as evidence of de-anchoring in long-run inflation expectations. However, if long-run expectations are persistently off target, such effects may instead reflect a re-anchoring channel, whereby policy deliberately shifts expectations back toward the target \citep[for detailed discussion, see, e.g., ][]{diegel2021long}. Consistent with this interpretation, our counterfactual analysis (panel (b) of Figure \ref{fig:histdecomp}) indicates that monetary policy shocks pulled trend inflation down and kept it closer to the 2\% target from the mid-1990s through the onset of the COVID-19 pandemic.

 
Taken together, our results suggest that central banks face a dilemma: bringing inflation back to target may come at the cost of lower potential GDP growth and a lower neutral interest rate. This, in turn, points to a potential role for fiscal interventions that support business investment and the economy's productive capacity during disinflation episodes.

\subsection{Robustness Analysis}\label{sec:Robustness checks}
We assess the robustness of our baseline findings by re-estimating the model under a range of alternative specifications. Figures \ref{fig:IRFtrends_estimatekappa}-\ref{fig:IRFtrends_EBP} show results of the IRFs of the trends and Figures \ref{fig:Historical_estimatekappa}-\ref{fig:Historical_EBP} show the historical decomposition of all alternative specifications. Across all variations, the qualitative responses of the macroeconomic stars to a contractionary monetary policy shock remain similar to those in the baseline, indicating that our main results are not driven by specific modelling choices.

First, estimating the shrinkage parameter $V_{b}$ via a hierarchical Bayes approach yields posterior impulse responses that closely match the baseline, with slightly wider credible intervals (Figure \ref{fig:IRFtrends_estimatekappa}) and very similar historical decompositions (Figure \ref{fig:Historical_estimatekappa}). Second, restricting the sample to pre-COVID observations ending in 2019Q4 produces similar negative responses of $g^{*}$, $\pi^{*}$, and $r^{*}$ (Figure \ref{fig:IRFtrends_excludeCovid}) as well as a similar counterfactual path of the trends (Figure \ref{fig:Historical_excludeCovid}), suggesting that the pandemic period does not drive the results. Third, increasing the VAR lag length from four to eight (Figures \ref{fig:IRFtrends_P=8} and \ref{fig:Historical_P=8}) does not materially alter the estimated responses or historical decompositions. Fourth, replacing the federal funds rate with the one-year or two-year Treasury yield (Figures \ref{fig:IRFtrends_GS1}, \ref{fig:IRFtrends_GS2}, \ref{fig:Historical_GS1} and \ref{fig:Historical_GS2}) preserves the qualitative pattern of the responses and the pattern of the counterfactual paths. Fifth, substituting the GDP deflator with the PCE price index (Figures \ref{fig:IRFtrends_PCEPI} and \ref{fig:Historical_PCEPI}) yields comparable estimates. Sixth, using the unadjusted monetary policy surprise measure from Bauer and Swanson (2023) instead of the orthogonalized version (Figures \ref{fig:IRFtrends_MPS} and \ref{fig:Historical_MPS}) results in nearly identical posterior distributions. Finally, controlling for financial conditions by adding the excess bond premium \citep{favara2016updating} to the cycle part of the model confirms our main empirical findings( Figures \ref{fig:IRFtrends_EBP} and \ref{fig:Historical_EBP}). 

Overall, these robustness analyses confirm that the estimated  effects of monetary policy shocks on the macroeconomic stars are stable across alternative sample periods, lag specifications, nominal interest rate measures, inflation measures, external instruments, and controlling for financial conditions using the excess bond premium.

%
%
%
%
	%
%
\vspace{-10pt}
	
	\section{Conclusions}

This paper develops SMUC-IV to explore the effects of monetary policy shocks on the macroeconomic stars, measured as long-run trends, in a unified framework. Using our SMUC-IV, we show that monetary tightening can have negative effects on the potential GDP growth, trend inflation, and the neutral interest rate, and that monetary policy is an important driver of these stars. From a policy perspective, our results indicate that policymakers can bring elevated inflation back to target through monetary tightening, but potentially at the cost of long-run damage to the real economy. This dilemma points to a role for complementary fiscal interventions that support business investment and the economy's productive capacity during disinflation episodes.  Recently, growing attention has been paid to the long-run effects not only of monetary policy shocks but also of other structural shocks, such as financial, fiscal, and demand shocks \citep{cerra2005did,cerra2008growth,antolin2025long,furlanetto2025estimating}. Against this backdrop, future research could extend our framework to examine how different types of structural shocks shape long-run macroeconomic equilibria.


	\small{\setstretch{0.85}
		\addcontentsline{toc}{section}{References}

		\bibliography{Lit}

@article{geweke1999using,
	title={{Using Simulation Methods for Bayesian Econometric Models: Inference, Development, and Communication}},
	author={Geweke, John},
	journal={Econometric reviews},
	volume={18},
	number={1},
	pages={1--73},
	year={1999},
	publisher={Taylor \& Francis}
}

@Article{stockwatson2,
  author  = {Stock, J. and Watson, M.},
  title   = {{Why has U.S. Inflation become Harder to Forecast?}},
  journal = {Journal of Money, Credit and Banking},
  year    = {2007},
  volume  = {39},
  number  = {7},
  pages   = {3-33},
}

@article{rue2001fast,
	title={{Fast Sampling of Gaussian Markov Random Fields}},
	author={Rue, H{\aa}vard},
	journal={Journal of the Royal Statistical Society: Series B (Statistical Methodology)},
	volume={63},
	number={2},
	pages={325--338},
	year={2001},
	publisher={Wiley Online Library}
}

@article{mccausland2011simulation,
	title={{Simulation Smoothing for State-Space Models: A Computational Efficiency Analysis}},
	author={McCausland, William J and Miller, Shirley and Pelletier, Denis},
	journal={Computational Statistics and Data Analysis},
	volume={55},
	number={1},
	pages={199--212},
	year={2011},
	publisher={Elsevier}
}

@article{miranda2021transmission,
	title={{The Transmission of Monetary Policy Shocks}},
	author={Miranda-Agrippino, Silvia and Ricco, Giovanni},
	journal={American Economic Journal: Macroeconomics},
	volume={13},
	number={3},
	pages={74--107},
	year={2021},
	publisher={American Economic Association 2014 Broadway, Suite 305, Nashville, TN 37203-2425}
}

@Article{chan2009efficient,
  author    = {Chan, Joshua CC and Jeliazkov, Ivan},
  journal   = {International Journal of Mathematical Modelling and Numerical Optimisation},
  title     = {{Efficient Simulation and Integrated Likelihood Estimation in State Space Models}},
  year      = {2009},
  number    = {1-2},
  pages     = {101--120},
  volume    = {1},
  publisher = {Inderscience Publishers},
}

@Article{chan2021minnesota,
  author    = {Chan, Joshua CC},
  journal   = {International Journal of Forecasting},
  title     = {{Minnesota-Type Adaptive Hierarchical Priors for Large Bayesian VARs}},
  year      = {2021},
  number    = {3},
  pages     = {1212--1226},
  volume    = {37},
  publisher = {Elsevier},
}

@Book{kilian2017structural,
  author    = {Kilian, Lutz and L{\"u}tkepohl, Helmut},
  publisher = {Cambridge University Press},
  title     = {{Structural Vector Autoregressive Analysis}},
  year      = {2017},
}

@Article{mertens2013dynamic,
  author    = {Mertens, Karel and Ravn, Morten O},
  journal   = {American Economic Review},
  title     = {The Dynamic Effects of Personal and Corporate Income Tax Changes in the United States},
  year      = {2013},
  number    = {4},
  pages     = {1212--1247},
  volume    = {103},
  publisher = {American Economic Association},
}

@Article{cross2020macroeconomic,
  author    = {Cross, Jamie L and Hou, Chenghan and Poon, Aubrey},
  journal   = {International Journal of Forecasting},
  title     = {{Macroeconomic Forecasting with Large Bayesian VARs: Global-Local Priors and the Illusion of Sparsity}},
  year      = {2020},
  number    = {3},
  pages     = {899--915},
  volume    = {36},
  publisher = {Elsevier},
}

@Article{braun2023identification,
  author    = {Braun, Robin and Br{\"u}ggemann, Ralf},
  journal   = {Journal of Business and Economic Statistics},
  title     = {{Identification of SVAR Models by Combining Sign Restrictions with External Instruments}},
  year      = {2023},
  number    = {4},
  pages     = {1077--1089},
  volume    = {41},
  publisher = {Taylor \& Francis},
}

@Article{hou2024large,
  author    = {Hou, Chenghan},
  journal   = {Journal of Econometrics},
  title     = {{Large Bayesian SVARs with Linear Restrictions}},
  year      = {2024},
  number    = {1},
  pages     = {105850},
  volume    = {244},
  publisher = {Elsevier},
}

@Article{jorda2024long,
  author    = {Jord{\`a}, {\`O}scar and Singh, Sanjay R and Taylor, Alan M},
  journal   = {Review of Economics and Statistics},
  title     = {{The Long-Run Effects of Monetary Policy}},
  year      = {2024},
  pages     = {1--49},
  publisher = {MIT Press 255 Main Street, 9th Floor, Cambridge, Massachusetts 02142, USA~},
}

@Article{del2017safety,
  author    = {Del Negro, Marco and Giannone, Domenico and Giannoni, Marc P and Tambalotti, Andrea},
  journal   = {Brookings Papers on Economic Activity},
  title     = {{Safety, Liquidity, and the Natural Rate of Interest}},
  year      = {2017},
  number    = {1},
  pages     = {235--316},
  volume    = {2017},
  publisher = {Johns Hopkins University Press},
}

@Article{fornaro2023scars,
	title={{The Scars of Supply Shocks: Implications for Monetary Policy}},
	author={Fornaro, Luca and Wolf, Martin},
	journal={Journal of Monetary Economics},
	volume={140},
	pages={S18--S36},
	year={2023},
	publisher={Elsevier} }

@Article{diegel2021long,
	title={{Long-Term Inflation Expectations and the Transmission of Monetary Policy Shocks: Evidence from a SVAR Analysis}},
	author={Diegel, Max and Nautz, Dieter},
	journal={Journal of Economic Dynamics and Control},
	volume={130},
	pages={104192},
	year={2021},
	publisher={Elsevier}
}

@Article{maffei2025identifying,
	title={{Identifying the Sources of the Slowdown in Growth: Demand versus Supply}},
	author={Maffei-Faccioli, Nicol{\`o}},
	journal={Journal of Applied Econometrics},
	volume={40},
	number={2},
	pages={181--194},
	year={2025},
	publisher={Wiley Online Library}
}

@Article{morley2003beveridge,
	title={{Why are the Beveridge-Nelson and Unobserved-Components Decompositions of GDP so Different?}},
	author={Morley, James C and Nelson, Charles R and Zivot, Eric},
	journal={Review of Economics and Statistics},
	volume={85},
	number={2},
	pages={235--243},
	year={2003},
	publisher={MIT Press 238 Main St., Suite 500, Cambridge, MA 02142-1046, USA journals~?}
}

@Article{basistha2007new,
	title={{New Measures of the Output Gap based on the Forward-Looking New Keynesian Phillips Curve}},
	author={Basistha, Arabinda and Nelson, Charles R},
	journal={Journal of Monetary Economics},
	volume={54},
	number={2},
	pages={498--511},
	year={2007},
	publisher={Elsevier}
}

@Article{grant2017reconciling,
	title={{Reconciling Output Gaps: Unobserved Components Model and Hodrick--Prescott Filter}},
	author={Grant, Angelia L and Chan, Joshua CC},
	journal={Journal of Economic Dynamics and Control},
	volume={75},
	pages={114--121},
	year={2017},
	publisher={Elsevier}
}

@Article{grant2017bayesian,
  author    = {Grant, Angelia L and Chan, Joshua CC},
  journal   = {Journal of Money, Credit and Banking},
  title     = {{A Bayesian Model Comparison for Trend-Cycle Decompositions of Output}},
  year      = {2017},
  number    = {2-3},
  pages     = {525--552},
  volume    = {49},
  publisher = {Wiley Online Library},
}

@Article{chan2013new,
	title={{A New Model of Trend Inflation}},
	author={Chan, Joshua CC and Koop, Gary and Potter, Simon M},
	journal={Journal of Business and Economic Statistics},
	volume={31},
	number={1},
	pages={94--106},
	year={2013},
	publisher={Taylor \& Francis}
}

@Article{chan2018new,
	title={{A New Model of Inflation, Trend Inflation, and Long-Run Inflation Expectations}},
	author={Chan, Joshua CC and Clark, Todd E and Koop, Gary},
	journal={Journal of Money, Credit and Banking},
	volume={50},
	number={1},
	pages={5--53},
	year={2018},
	publisher={Wiley Online Library}
}

@Article{mertens2016measuring,
	title={{Measuring the Level and Uncertainty of Trend Inflation}},
	author={Mertens, Elmar},
	journal={Review of Economics and Statistics},
	volume={98},
	number={5},
	pages={950--967},
	year={2016},
	publisher={The MIT Press}
}

@article{stock2016core,
	title={{Core Inflation and Trend Inflation}},
	author={Stock, James H and Watson, Mark W},
	journal={Review of Economics and Statistics},
	volume={98},
	number={4},
	pages={770--784},
	year={2016},
	publisher={The MIT Press}
}

@Article{hwu2019estimating,
	title={{Estimating Trend Inflation Based on Unobserved Components Model: Is It Correlated with the Inflation Gap?}},
	author={Hwu, Shih-tang and Kim, Chang-jin},
	journal={Journal of Money, Credit and Banking},
	volume={51},
	number={8},
	pages={2305--2319},
	year={2019},
	publisher={Wiley Online Library}
}

@Article{laubach2003measuring,
	title={{Measuring the Natural Rate of Interest}},
	author={Laubach, Thomas and Williams, John C},
	journal={Review of Economics and Statistics},
	volume={85},
	number={4},
	pages={1063--1070},
	year={2003},
	publisher={MIT Press 238 Main St., Suite 500, Cambridge, MA 02142-1046, USA journals~?}
}

@Article{holston2017measuring,
	title={{Measuring the Natural Rate of Interest: International Trends and Determinants}},
	author={Holston, Kathryn and Laubach, Thomas and Williams, John C},
	journal={Journal of International Economics},
	volume={108},
	pages={S59--S75},
	year={2017},
	publisher={Elsevier}
}

@Article{zaman2025unified,
	title={{A Unified Framework to Estimate Macroeconomic Stars}},
	author={Zaman, Saeed},
	journal={Review of Economics and Statistics},
	pages={1--45},
	year={2025},
	publisher={MIT Press 255 Main Street, 9th Floor, Cambridge, Massachusetts 02142, USA~?}
}

@Article{antolin2025long,
	title={{The Long-Run Effects of Government Spending}},
	author={Antolin-Diaz, Juan and Surico, Paolo},
	journal={American Economic Review},
	volume={115},
	number={7},
	pages={2376--2413},
	year={2025},
	publisher={American Economic Association 2014 Broadway, Suite 305, Nashville, TN 37203}
}

@Article{furlanetto2025estimating,
	title={{Estimating Hysteresis Effects}},
	author={Furlanetto, Francesco and Lepetit, Antoine and Robstad, {\O}rjan and Rubio-Ram{\'\i}rez, Juan and Ulvedal, P{\aa}l},
	journal={American Economic Journal: Macroeconomics},
	volume={17},
	number={1},
	pages={35--70},
	year={2025},
	publisher={American Economic Association 2014 Broadway, Suite 305, Nashville, TN 37203-2425}
}

@article{cerra2005did,
	title={{Did Output Recover from the Asian Crisis?}},
	author={Cerra, Valerie and Saxena, Sweta Chaman},
	journal={IMF Staff Papers},
	volume={52},
	number={1},
	pages={1--23},
	year={2005},
	publisher={Springer}
}

@article{cerra2008growth,
	title={{Growth Dynamics: the Myth of Economic Recovery}},
	author={Cerra, Valerie and Saxena, Sweta Chaman},
	journal={American Economic Review},
	volume={98},
	number={1},
	pages={439--457},
	year={2008},
	publisher={American Economic Association}
}

@article{cerra2023hysteresis,
	title={{Hysteresis and Business Cycles}},
	author={Cerra, Valerie and Fat{\'a}s, Antonio and Saxena, Sweta C},
	journal={Journal of Economic Literature},
	volume={61},
	number={1},
	pages={181--225},
	year={2023},
	publisher={American Economic Association 2014 Broadway, Suite 305, Nashville, TN 37203-2425}
}

@article{garga2021output,
	title={Output hysteresis and optimal monetary policy},
	author={Garga, Vaishali and Singh, Sanjay R},
	journal={Journal of Monetary Economics},
	volume={117},
	pages={871--886},
	year={2021},
	publisher={Elsevier}
}

@article{moran2018innovation,
	title={{Innovation, Productivity, and Monetary Policy}},
	author={Moran, Patrick and Queralto, Albert},
	journal={Journal of Monetary Economics},
	volume={93},
	pages={24--41},
	year={2018},
	publisher={Elsevier}
}

@article{watson1986univariate,
	title={{Univariate Detrending Methods with Stochastic Trends}},
	author={Watson, Mark W},
	journal={Journal of Monetary Economics},
	volume={18},
	number={1},
	pages={49--75},
	year={1986},
	publisher={Elsevier}
}

@article{eo2023understanding,
	title={{Understanding Trend Inflation through the Lens of the Goods and Services Sectors}},
	author={Eo, Yunjong and Uzeda, Luis and Wong, Benjamin},
	journal={Journal of Applied Econometrics},
	volume={38},
	number={5},
	pages={751--766},
	year={2023},
	publisher={Wiley Online Library}
}

@Article{lubik2015calculating,
  author    = {Lubik, Thomas A and Matthes, Christian},
  journal   = {Richmond Fed Economic Brief},
  title     = {{Calculating the Natural Rate of Interest: A Comparison of Two Alternative Approaches}},
  year      = {2015},
  publisher = {Federal Reserve Bank of Richmond},
}

@article{morley2024simple,
	title={{A Simple Correction for Misspecification in Trend-Cycle Decompositions with an Application to Estimating R}},
	author={Morley, James and Tran, Trung Duc and Wong, Benjamin},
	journal={Journal of Business and Economic Statistics},
	volume={42},
	number={2},
	pages={665--680},
	year={2024},
	publisher={Taylor \& Francis}
}

@article{beveridge1981new,
	title={{A New Approach to Decomposition of Economic Time Series into Permanent and Transitory Components with Particular Attention to Measurement of the `Business Cycle'}},
	author={Beveridge, Stephen and Nelson, Charles R},
	journal={Journal of Monetary Economics},
	volume={7},
	number={2},
	pages={151--174},
	year={1981},
	publisher={Elsevier}
}

@techreport{ma2023monetary,
	title={{Monetary Policy and Innovation}},
	author={Ma, Yueran and Zimmermann, Kaspar},
	year={2023},
	institution={National Bureau of Economic Research}
}

@article{meier2024monetary,
	title={{Monetary Policy, Markup Dispersion, and Aggregate TFP}},
	author={Meier, Matthias and Reinelt, Timo},
	journal={Review of Economics and Statistics},
	volume={106},
	number={4},
	pages={1012--1027},
	year={2024},
	publisher={MIT Press 255 Main Street, 9th Floor, Cambridge, Massachusetts 02142, USA~?}
}

@article{stock2018identification,
	title={{Identification and Estimation of Dynamic Causal Effects in Macroeconomics Using External Instruments}},
	author={Stock, James H and Watson, Mark W},
	journal={The Economic Journal},
	volume={128},
	number={610},
	pages={917--948},
	year={2018},
	publisher={Oxford University Press Oxford, UK}
}

@article{arias2021inference,
	title={{Inference in Bayesian Proxy-SVARs}},
	author={Arias, Jonas E and Rubio-Ramirez, Juan F and Waggoner, Daniel F},
	journal={Journal of Econometrics},
	volume={225},
	number={1},
	pages={88--106},
	year={2021},
	publisher={Elsevier}
}

@article{caldara2019monetary,
	title={{Monetary Policy, Real Activity, and Credit Spreads: Evidence from Bayesian Proxy SVARs}},
	author={Caldara, Dario and Herbst, Edward},
	journal={American Economic Journal: Macroeconomics},
	volume={11},
	number={1},
	pages={157--192},
	year={2019},
	publisher={American Economic Association 2014 Broadway, Suite 305, Nashville, TN 37203-2425}
}

@article{wu2016measuring,
	title={{Measuring the Macroeconomic Impact of Monetary Policy at the Zero Lower Bound}},
	author={Wu, Jing Cynthia and Xia, Fan Dora},
	journal={Journal of Money, Credit and Banking},
	volume={48},
	number={2-3},
	pages={253--291},
	year={2016},
	publisher={Wiley Online Library}
}

@article{bauer2023reassessment,
	title={{A Reassessment of Monetary Policy Surprises and High-Frequency Identification}},
	author={Bauer, Michael D and Swanson, Eric T},
	journal={NBER Macroeconomics Annual},
	volume={37},
	number={1},
	pages={87--155},
	year={2023},
	publisher={The University of Chicago Press Chicago, IL}
}

@Article{favara2016updating,
  author  = {Favara, Giovanni and Gilchrist, Simon and Lewis, Kurt F and Zakraj{\v{s}}ek, Egon},
  journal = {FEDS Notes},
  title   = {{Updating the Recession Risk and the Excess Bond Premium}},
  year    = {2016},
}

@article{leiva2023endogenous,
	title={{Endogenous Time Variation in Vector Autoregressions}},
	author={Leiva-Le{\'o}n, Danilo and Uzeda, Luis},
	journal={Review of Economics and Statistics},
	volume={105},
	number={1},
	pages={125--142},
	year={2023},
	publisher={MIT Press One Rogers Street, Cambridge, MA 02142-1209, USA journals-info~?}
}

@article{jarocinski2020deconstructing,
	title={{Deconstructing Monetary Policy Surprises? --- The Role of Information Shocks}},
	author={Jaroci{\'n}ski, Marek and Karadi, Peter},
	journal={American Economic Journal: Macroeconomics},
	volume={12},
	number={2},
	pages={1--43},
	year={2020},
	publisher={American Economic Association 2014 Broadway, Suite 305, Nashville, TN 37203-2425}
}

@article{kuttner1994estimating,
	title={{Estimating Potential Output as A Latent Variable}},
	author={Kuttner, Kenneth N},
	journal={Journal of Business and Economic Statistics},
	volume={12},
	number={3},
	pages={361--368},
	year={1994},
	publisher={Taylor \& Francis}
}

@article{chan2016bounded,
	title={{A Bounded Model of Time Variation in Trend Inflation, NAIRU and the Phillips Curve}},
	author={Chan, Joshua CC and Koop, Gary and Potter, Simon M},
	journal={Journal of Applied Econometrics},
	volume={31},
	number={3},
	pages={551--565},
	year={2016},
	publisher={Wiley Online Library}
}

@article{ramey2016macroeconomic,
	title={{Macroeconomic shocks and their propagation}},
	author={Ramey, Valerie A},
	journal={Handbook of Macroeconomics},
	volume={2},
	pages={71--162},
	year={2016},
	publisher={Elsevier}
}

@article{gelfand1994bayesian,
	title={{Bayesian Model Choice: Asymptotics and Exact Calculations}},
	author={Gelfand, Alan E and Dey, Dipak K},
	journal={Journal of the Royal Statistical Society: Series B (Methodological)},
	volume={56},
	number={3},
	pages={501--514},
	year={1994},
	publisher={Wiley Online Library}
}

@article{chan2023comparing,
	title={{Comparing Stochastic Volatility Specifications for Large Bayesian VARs}},
	author={Chan, Joshua CC},
	journal={Journal of Econometrics},
	volume={235},
	number={2},
	pages={1419--1446},
	year={2023},
	publisher={Elsevier}
}
		\appendix
		
		\renewcommand{\thesection}{Appendix \Alph{section}}
		\setcounter{table}{0}
		\renewcommand{\thetable}{\Alph{section}.\arabic{table}}
		\setcounter{figure}{0}
		\renewcommand{\thefigure}{\Alph{section}.\arabic{figure}}
		
		\newpage
		\clearpage

		\section{Estimation Details}

This Appendix provides details of Step 3 - Step 5 of the posterior sampler. 

To sample $\boldsymbol{\Phi}$ in Step 3, we first rewrite the model in \eqref{eq:aug_SVAR} as
\begin{align}
	\bar{\boldsymbol{\eta}}_t = \bar{\mathbf{A}}_1 \bar{\boldsymbol{\eta}}_{t-1} + \ldots + \bar{\mathbf{A}}_p \bar{\boldsymbol{\eta}}_{t-p} + \widetilde{\mathbf{u}}_t, \quad \widetilde{\mathbf{u}}_t \sim \mathcal{N}(\mathbf{0}, \widetilde{\boldsymbol{\Sigma}}), \label{eq:VAR_phi}
\end{align}
where $\bar{\boldsymbol{\eta}}_t = (g_t^{*} - g_{t-1}^{*}, \pi_t^{*} - \pi_{t-1}^{*}, r_t^{*} - r_{t-1}^{*}, \mathbf{c}_t', m_t)'$, $\widetilde{\boldsymbol{\Sigma}} = \widetilde{\mathbf{B}} \widetilde{\mathbf{B}}'$ and $\bar{\mathbf{A}}_i = \text{diag}(\mathbf{0}_3, \boldsymbol{\Phi}_i, 0)$ for $i = 1, \ldots, p$. Stacking equation~\eqref{eq:VAR_phi} over $t = 1, \ldots, T$, we have
\begin{align}
	\bar{\boldsymbol{\eta}} = \bar{\mathbf{X}} \bar{\mathbf{a}} + \widetilde{\mathbf{u}}, \quad \widetilde{\mathbf{u}} \sim \mathcal{N}(\mathbf{0}, \mathbf{I}_T \otimes \widetilde{\boldsymbol{\Sigma}}), \label{eq:compact_eta}
\end{align}
where $\bar{\boldsymbol{\eta}} = (\bar{\boldsymbol{\eta}}_1', \ldots, \bar{\boldsymbol{\eta}}_T')'$, $\bar{\mathbf{a}} = \text{vec}\left( (\bar{\mathbf{A}}_1, \ldots, \bar{\mathbf{A}}_p)' \right)$, $\bar{\mathbf{X}} = ( \bar{\mathbf{X}}_1', \ldots, \bar{\mathbf{X}}_T')'$ with $\bar{\mathbf{X}}_t = \mathbf{I}_7 \otimes (\bar{\boldsymbol{\eta}}_{t-1}', \ldots, \bar{\boldsymbol{\eta}}_{t-p}')$. Let $\boldsymbol{\phi} = \text{vec}\left( (\boldsymbol{\Phi}_1, \ldots, \boldsymbol{\Phi}_p)'\right) $ and $\mathbf{S}_{\phi}$, then there is a selection matrix $\mathbf{S}_{\phi}$ such that $\bar{\mathbf{a}} = \mathbf{S}_{\phi} \boldsymbol{\phi}$. It follows that we can write \eqref{eq:compact_eta} as 
\begin{align*}
	\bar{\boldsymbol{\eta}} = \mathbf{X} \boldsymbol{\phi} + \widetilde{\mathbf{u}}, \quad \widetilde{\mathbf{u}} \sim \mathcal{N}(\mathbf{0}, \mathbf{I}_T \otimes \widetilde{\boldsymbol{\Sigma}}),
\end{align*}
where $\mathbf{X} = \bar{\mathbf{X}} \mathbf{S}_{\phi}$. Since we consider a Gaussian prior $\boldsymbol{\phi} \sim \mathcal{N}(\boldsymbol{\phi}_0, \mathbf{V}_{\phi_0})$, where the prior mean $\boldsymbol{\phi}_0$ and the diagonal covariance matrix $\mathbf{V}_{\phi_0}$ can be constructed as described in Section~\ref{sec:Priors}. Using the standard Bayesian linear regression results, we can obtain the conditional posterior
\begin{align*}
	( \boldsymbol{\phi}| \boldsymbol{\tau}, \mathbf{B}, \beta, \alpha, \kappa_{1}, \kappa_{2}, \mathbf{y}) \sim \mathcal{N}(\widehat{\boldsymbol{\phi}}, \widehat{\mathbf{V}}_{\phi} ),
\end{align*}
where $\widehat{\mathbf{V}}_{\phi} = ( \mathbf{X}' (\mathbf{I}_T \otimes \widetilde{\boldsymbol{\Sigma}}) \mathbf{X} + \mathbf{V}_{\phi_0}^{-1} )^{-1}$ and $\widehat{\boldsymbol{\phi}} = \widehat{\mathbf{V}}_{\phi} ( \mathbf{X}' (\mathbf{I}_T \otimes \widetilde{\boldsymbol{\Sigma}}) \bar{\boldsymbol{\eta}} + \mathbf{V}_{\phi_0}^{-1} \boldsymbol{\phi}_0  )$.

To obtain the conditional posteriors of $\kappa_1$ and $\kappa_2$ in Step 4 and Step 5, we first define 
\begin{align*}
	\widetilde{\boldsymbol{\Phi}}_{l,i,j} &= 
	\begin{dcases}
		l^2  \left( \boldsymbol{\Phi}_{l,i,j}  - \phi_{l,i,j}   \right)^2, \quad \text{$l = 1, \ldots, p$, $i, j = 1, 2, 3$, $i = j$}, \\
		\frac{\sigma_j^2}{\sigma_i^2} l^2  \left( \boldsymbol{\Phi}_{l,i,j}  - \phi_{l,i,j}   \right)^2, \quad \text{$l = 1, \ldots, p$, $i, j = 1, 2, 3$, $i \neq j$},
	\end{dcases}
\end{align*}
Then, it can be shown that the density functions of the conditional posteriors of $\kappa_1$ and $\kappa_2$ are given by
\begin{align*}
	&p(\kappa_1 | \boldsymbol{\tau}, \mathbf{B}, \beta, \alpha, \boldsymbol{\Phi}, \mathbf{y}) \propto \kappa_1^{-\frac{3 p}{2}} e^{- \frac{1}{ 2 \kappa_1} \sum_{i=j} \sum_{l=1}^{p} \widetilde{\boldsymbol{\Phi}}_{l,i,j} } \mathbf{1}( 0 < \kappa_1 < 1), \\
	&p(\kappa_2 | \boldsymbol{\tau}, \mathbf{B}, \beta, \alpha, \boldsymbol{\Phi}, \mathbf{y}) \propto \kappa_2^{-\frac{6 p}{2}} e^{- \frac{1}{ 2 \kappa_2} \sum_{i \neq j}  \sum_{l=1}^{p} \widetilde{\boldsymbol{\Phi}}_{l,i,j} } \mathbf{1}( 0 < \kappa_2 < 1).
\end{align*}
This implies that the conditional posteriors of $\kappa_1$ and $\kappa_2$ are truncated inverse-Gamma distributions:
\begin{align*}
	(\kappa_1 | \boldsymbol{\tau}, \mathbf{B}, \beta, \alpha, \boldsymbol{\Phi}, \mathbf{y}) &\sim  \mathcal{IG}\left( \frac{3 p}{2} - 1, \frac{1}{2} \sum_{i=j} \sum_{l=1}^{p} \widetilde{\boldsymbol{\Phi}}_{l,i,j}  \right) \mathbf{1}( 0 < \kappa_1 < 1), \\
	(\kappa_2 | \boldsymbol{\tau}, \mathbf{B}, \beta, \alpha, \boldsymbol{\Phi}, \mathbf{y}) &\sim  \mathcal{IG}\left( \frac{6p}{2} - 1,  \frac{1}{2} \sum_{i \neq j}  \sum_{l=1}^{p} \widetilde{\boldsymbol{\Phi}}_{l,i,j} \right) \mathbf{1}( 0 < \kappa_2 < 1).
\end{align*}

		\section{Details of Marginal Likelihood Estimation}
In this Appendix, we first present the analytical expression of the marginal prior $p(\boldsymbol{\Phi})$ in \eqref{eq:ML_kappa}, and then detail the construction of the tuning function $q(\boldsymbol{\Phi}, \mathbf{B}, \alpha, \beta)$.
\subsection*{Expressions for $p(\boldsymbol{\Phi})$}
Given the priors of $\boldsymbol{\Phi}$ and $(\kappa_1, \kappa_2)$ described in Section~\ref{sec:Priors}, we have
\begin{align*}
	p(\boldsymbol{\Phi}) &= \int p(\boldsymbol{\Phi} | \kappa_1, \kappa_2) p(\kappa_1) p(\kappa_2) d(\kappa_1, \kappa_2)\\
	&= c_{\kappa} \int \kappa_1^{-\frac{3p}{2}} \kappa_2^{-\frac{6p}{2}} e^{- \frac{1}{ 2 \kappa_1} \sum_{i=j} \sum_{l=1}^{p} \widetilde{\boldsymbol{\Phi}}_{l,i,j} } e^{- \frac{1}{ 2 \kappa_2} \sum_{i \neq j}  \sum_{l=1}^{p} \widetilde{\boldsymbol{\Phi}}_{l,i,j} } \mathbf{1}(0 < \kappa_1 < 1, 0 < \kappa_2 < 1) d(\kappa_1, \kappa_2) \\
	&= c_{\kappa} \left( \int_{0}^1 \kappa_1^{-\frac{3p}{2}}  e^{- \frac{1}{ 2 \kappa_1} \sum_{i=j} \sum_{l=1}^{p} \widetilde{\boldsymbol{\Phi}}_{l,i,j} } d \kappa_1 \right) \times \left(  \int_{0}^1 \kappa_2^{-\frac{6p}{2}} e^{- \frac{1}{ 2 \kappa_2} \sum_{i \neq j}  \sum_{l=1}^{p} \widetilde{\boldsymbol{\Phi}}_{l,i,j} } d \kappa_2 \right)\\
	&= c_{\kappa} \times \Gamma(\hat{\nu}_1) \widehat{S}_1^{-\hat{\nu}_1} F_{\mathcal{IG}}(1; \hat{\nu}_1 \hat{S}_1) \times \Gamma(\hat{\nu}_2) \widehat{S}_2^{-\hat{\nu}_2} F_{\mathcal{IG}}(1; \hat{\nu}_2 \hat{S}_2),
\end{align*}
where the normalizing constant $c_{\kappa} = (2 \pi)^{-\frac{9p}{2}}  \prod_{i=1}^3 \prod_{j=1}^3 \prod_{l = 1}^p \frac{\sigma_j l}{\sigma_i}$, $\widetilde{\boldsymbol{\Phi}}_{l,i,j}$ is defined in Appendix~A, $F_{\mathcal{IG}}( x; \nu, S)$ denotes the cumulative distribution function of an inverse-gamma distribution with shape parameter $\nu$ and scale parameter $S$ evaluated at $x$, and
\begin{align*}
	\hat{\nu}_1= \frac{3p}{2} - 1, \quad \hat{S}_1 =  \frac{1}{ 2} \sum_{i=j} \sum_{l=1}^{p} \widetilde{\boldsymbol{\Phi}}_{l,i,j}, \quad \hat{\nu}_2 = \frac{6p}{2} - 1, \quad \hat{S}_2 =  \frac{1}{ 2 } \sum_{i \neq j}  \sum_{l=1}^{p} \widetilde{\boldsymbol{\Phi}}_{l,i,j}.
\end{align*}

\subsection*{Details about the Tuning Density Function}
This section provides the expressions for the truncated Gaussian densities $q_{\Phi}(\boldsymbol{\Phi})$, $q_B(\mathbf{B})$, $q_{\alpha}(\alpha)$, and $q_{\beta}(\beta)$, which are given as follows:
\begin{align*}
	q_{\Phi}(\boldsymbol{\Phi}) &= c^{-1}_{\Phi} (2 \pi)^{-\frac{9p}{2}} |\widehat{\boldsymbol{\Sigma}}_{\Phi}|^{-\frac{1}{2}} e^{-\frac{1}{2} \left( \text{vec}(\boldsymbol{\Phi}) - \text{vec}( \widehat{\boldsymbol{\Phi}}) \right)' \widehat{\boldsymbol{\Sigma}}_{\Phi}^{-1} \left( \text{vec}(\boldsymbol{\Phi}) - \text{vec}( \widehat{\boldsymbol{\Phi}}) \right)   } \mathbf{1}( \text{vec}(\boldsymbol{\Phi}) \in \mathcal{R}_{\Phi}),\\
	q_{B}(\mathbf{B}) &= c^{-1}_B (2 \pi)^{-18} |\widehat{\boldsymbol{\Sigma}}_{B}|^{-\frac{1}{2}} e^{-\frac{1}{2} \left( \text{vec}(\mathbf{B}) - \text{vec}( \widehat{\mathbf{B}}) \right)' \widehat{\boldsymbol{\Sigma}}_{B}^{-1} \left( \text{vec}(\mathbf{B}) - \text{vec}( \widehat{\mathbf{B}}) \right)   } \mathbf{1}( \text{vec}(\mathbf{B}) \in \mathcal{R}_{B}), \\
	q_{\beta}(\beta) &= c_{\beta}^{-1} (2 \pi \widehat{\sigma}^{2}_{\beta})^{-\frac{1}{2}} e^{-\frac{1}{2 \widehat{\sigma}^{2}_{\beta} } ( \beta - \widehat{\beta}  )^2  } \mathbf{1}(\beta \in \mathcal{R}_{\beta}) , \\
	q_{\alpha}(\alpha) &= c_{\alpha}^{-1} (2 \pi \widehat{\sigma}^{2}_{\alpha})^{-\frac{1}{2}} e^{-\frac{1}{2 \widehat{\sigma}^{2}_{\alpha} } ( \alpha - \widehat{\alpha}  )^2  } \mathbf{1}(\alpha \in \mathcal{R}_{\alpha}),
\end{align*}
where $c_{\Phi}$, $c_{B}$, $c_{\beta}$, and $c_{\alpha}$ are normalizing constants. The parameters $\widehat{\boldsymbol{\Phi}}$, $\widehat{\mathbf{B}}$, $\widehat{\beta}$, and $\widehat{\alpha}$ are the estimated posterior means, while $\widehat{\boldsymbol{\Sigma}}_{\Phi}$, $\widehat{\boldsymbol{\Sigma}}_{B}$, $\widehat{\sigma}^2_{\beta}$, and $\widehat{\sigma}^2_{\alpha}$ are their corresponding estimated covariance matrices and variances. The truncation regions are given by:
\begin{align*}
	\mathcal{R}_{\Phi} &= \left\{\boldsymbol{\Phi} :   \left( \text{vec}(\boldsymbol{\Phi}) - \text{vec}( \widehat{\boldsymbol{\Phi}}) \right)' \widehat{\boldsymbol{\Sigma}}_{\Phi}^{-1} \left( \text{vec}(\boldsymbol{\Phi}) - \text{vec}( \widehat{\boldsymbol{\Phi}}) \right) < \chi^2_{0.95, 9p} \right\}, \\
	\mathcal{R}_{B} &= \left\{ \mathbf{B} :   \left( \text{vec}(\mathbf{B}) - \text{vec}( \widehat{\mathbf{B}}) \right)' \widehat{\boldsymbol{\Sigma}}_{B}^{-1} \left( \text{vec}(\mathbf{B}) - \text{vec}( \widehat{\mathbf{B}}) \right) < \chi^2_{0.95, 36} \right\}, \\
	\mathcal{R}_{\beta} &= \left\{ \beta : \frac{(\beta - \widehat{\beta})^2}{\widehat{\sigma}^2_{\beta}  }  < \chi^2_{0.95, 1} \right\}, \\
	\mathcal{R}_{\alpha} &= (0, w ),
\end{align*}
where $\chi^2_{k_1,k_2}$ denotes the $k_1$ quantile of the chi-square distribution with $k_2$ degrees of freedom. The upper bound $w$ for $\mathcal{R}_{\alpha}$ is chosen such that the probability mass of the untruncated Gaussian $\mathcal{N}(\widehat{\alpha}, \widehat{\sigma}^2_{\alpha})$ over the interval $(0, w)$ is 0.95. Formally, $w$ satisfies:
\begin{align*}
	\mathcal{N}_{cdf} \left( \frac{w - \widehat{\alpha}}{ \widehat{\sigma}_{\alpha} } \right) - \mathcal{N}_{cdf} \left(- \frac{ \widehat{\alpha}  }{\widehat{\sigma}_{\alpha}  } \right) = 0.95,
\end{align*}
where $\mathcal{N}_{cdf}(\cdot)$ is the cumulative distribution function of the standard normal distribution.

		\section{Results Robustness Checks}

\begin{figure}[H]
	\centering
	\includegraphics[width=1\linewidth]{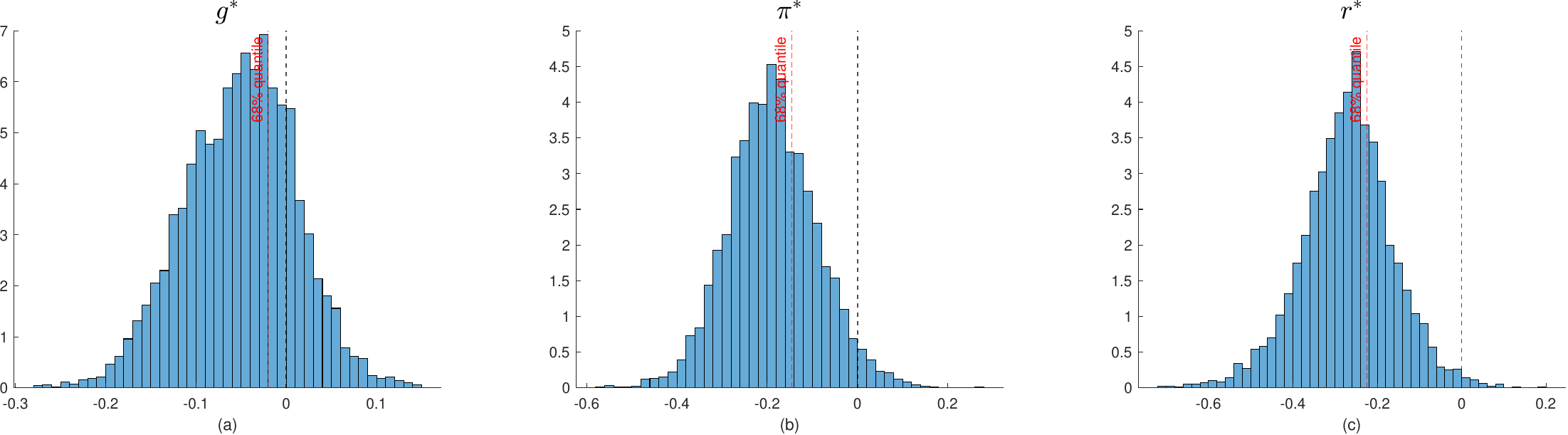}
	\caption{The plot shows the posterior distributions of the impulse response of the trends to a monetary policy shock. In this plot we estimate the shrinkage parameter 
	$V_{b} $ using a hierarchical Bayes approach. }
	\label{fig:IRFtrends_estimatekappa}
\end{figure}

\begin{figure}[H]
	\centering
	\includegraphics[width=1\linewidth]{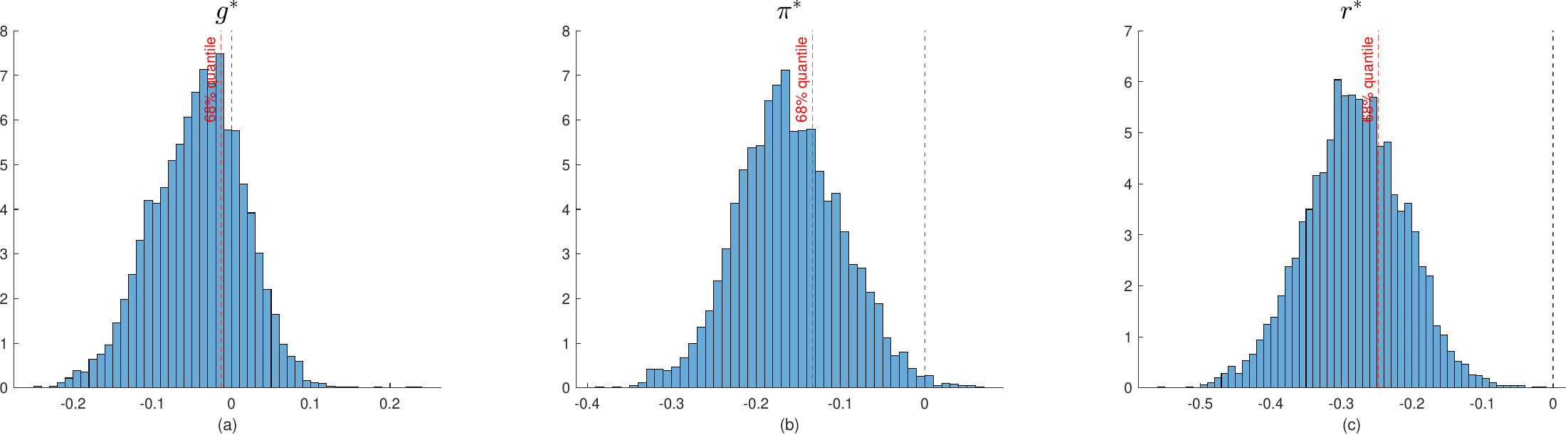}
	\caption{The plot shows the posterior distributions of the impulse response of the trends to a monetary policy shock. In this plot we use only data until 2019Q4. }
	\label{fig:IRFtrends_excludeCovid}
\end{figure}

i
\begin{figure}[H]
	\centering
	\includegraphics[width=1\linewidth]{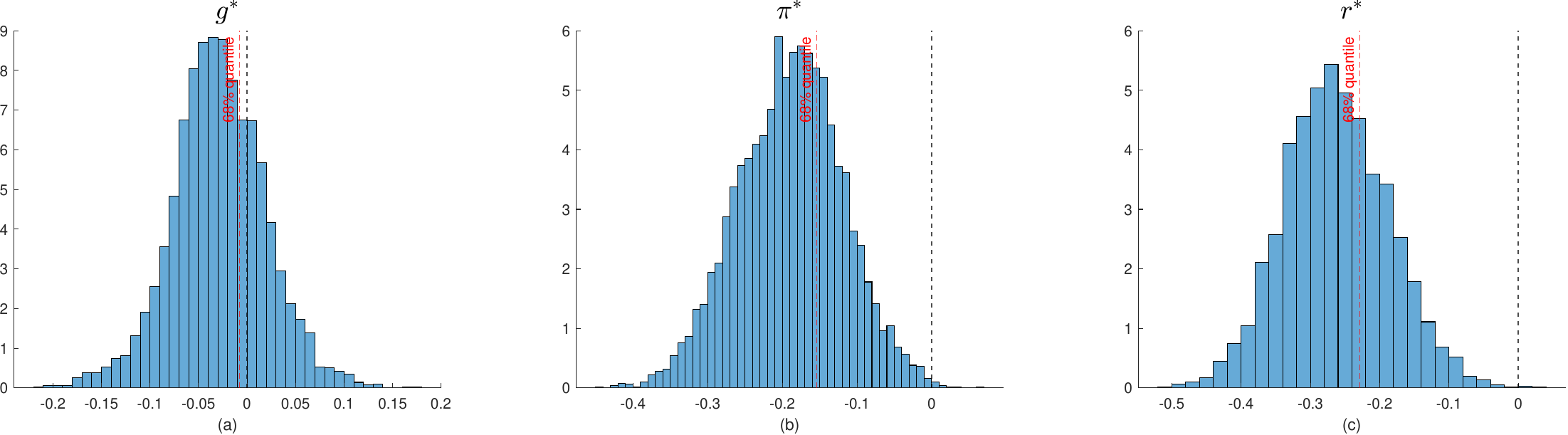}
	\caption{The plot shows the posterior distributions of the impulse response of the trends to a monetary policy shock. In this plot we estimate the model with eight lags instead of four. }
	\label{fig:IRFtrends_P=8}
\end{figure}

\begin{figure}[H]
	\centering
	\includegraphics[width=1\linewidth]{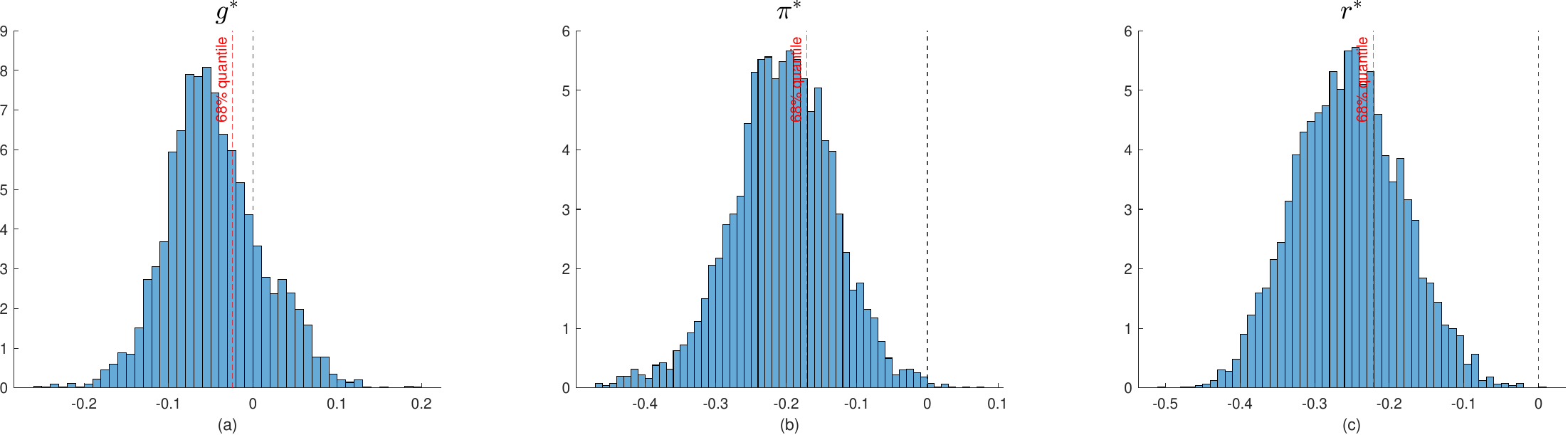}
	\caption{The plot shows the posterior distributions of the impulse response of the trends to a monetary policy shock. In this plot we estimate the model with the one-year treasury yield instead of using the federal funds effective rate. }
	\label{fig:IRFtrends_GS1}
\end{figure}

\begin{figure}[H]
	\centering
	\includegraphics[width=1\linewidth]{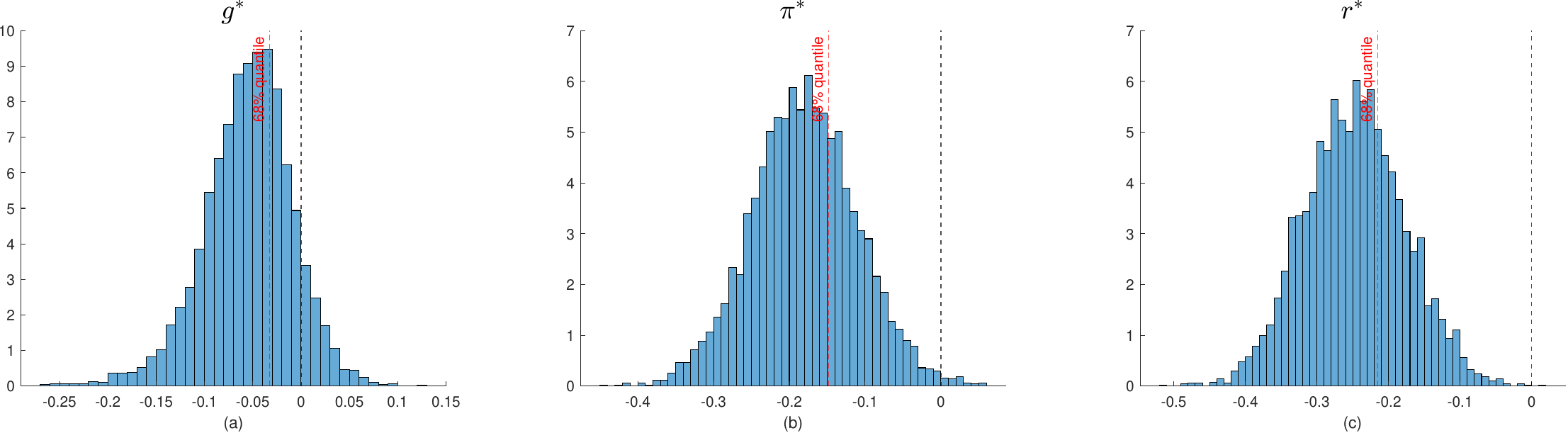}
	\caption{The plot shows the posterior distributions of the impulse response of the trends to a monetary policy shock. In this plot we estimate the model with the two-year treasury yield instead of using the federal funds effective rate. }
	\label{fig:IRFtrends_GS2}
\end{figure}

\begin{figure}[H]
	\centering
	\includegraphics[width=1\linewidth]{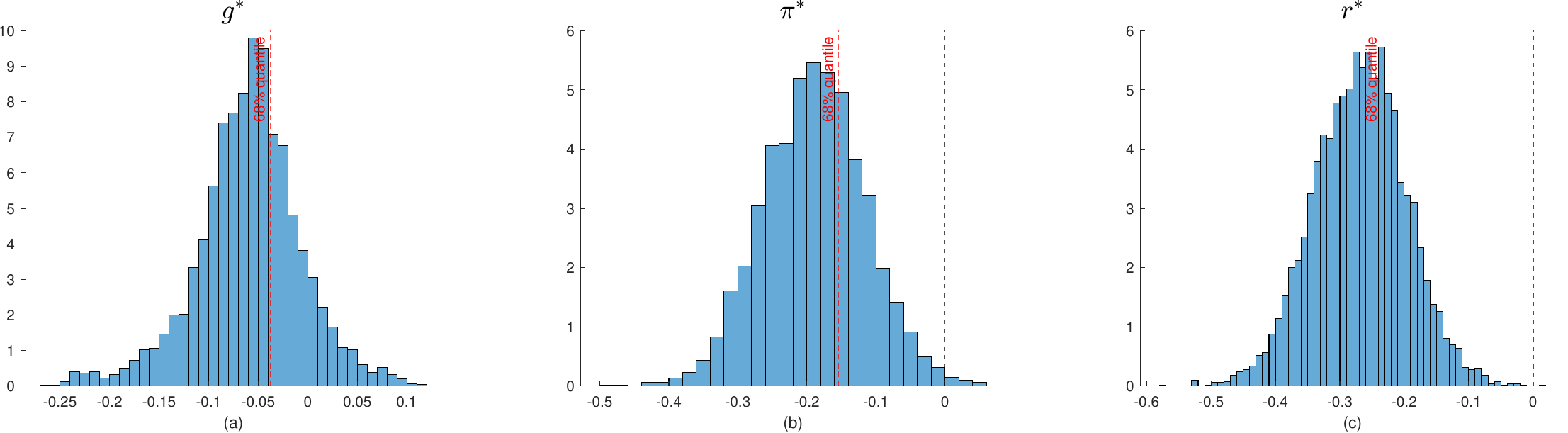}
	\caption{The plot shows the posterior distributions of the impulse response of the trends to a monetary policy shock. In this plot we estimate the model with the personal consumption expenditure price index instead of the GDP deflator. }
	\label{fig:IRFtrends_PCEPI}
\end{figure}

\begin{figure}[H]
	\centering
	\includegraphics[width=1\linewidth]{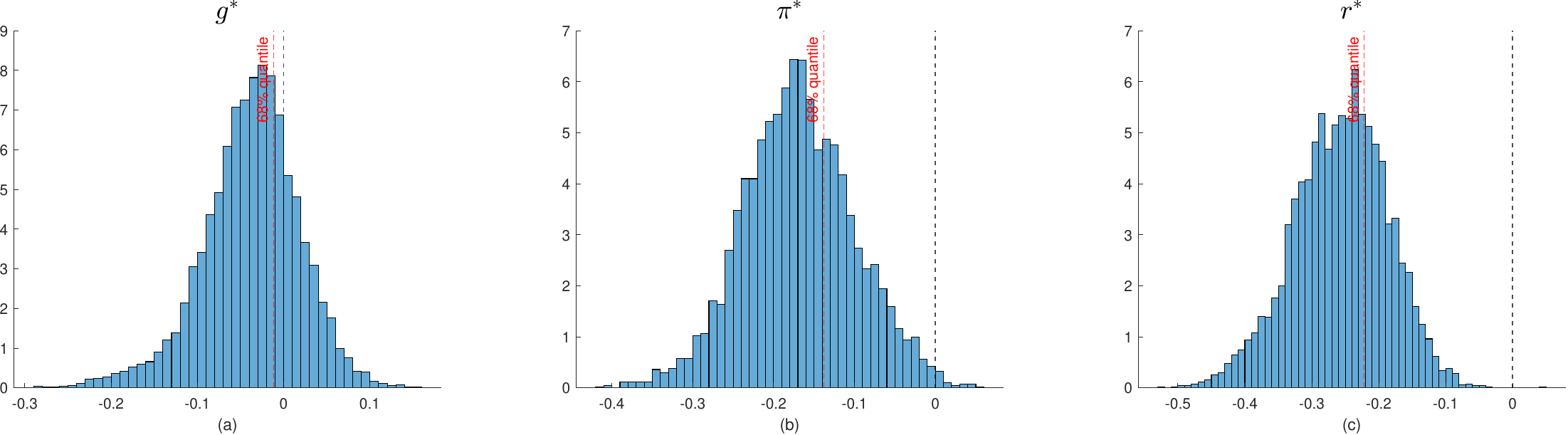}
	\caption{The plot shows the posterior distributions of the impulse response of the trends to a monetary policy shock. In this plot we estimate the model with unadjusted monetary policy surprise (MPS) measure from \citep{bauer2023reassessment} instead of using the orthogonalized monetary policy surprise measure. }
	\label{fig:IRFtrends_MPS}
\end{figure}

\begin{figure}[H]
	\centering
	\includegraphics[width=1\linewidth]{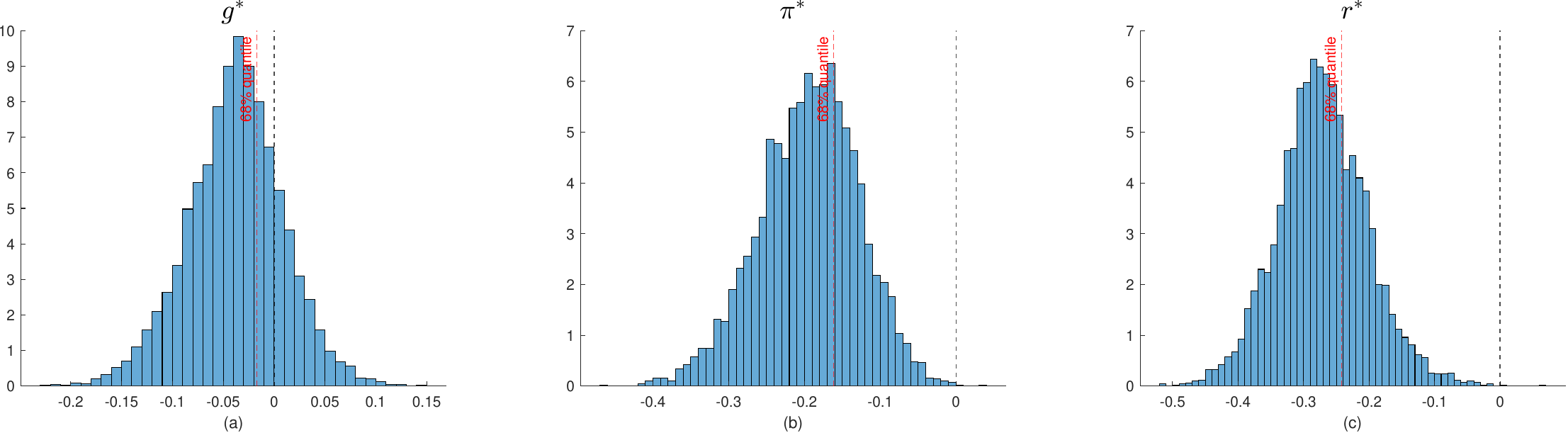}
	\caption{The plot shows the posterior distributions of the impulse response of the trends to a monetary policy shock. In this plot we estimate the model with adding the excess bond premium.  }
	\label{fig:IRFtrends_EBP}
\end{figure}

\begin{figure}[H]
	\centering
	\includegraphics[width=1\linewidth]{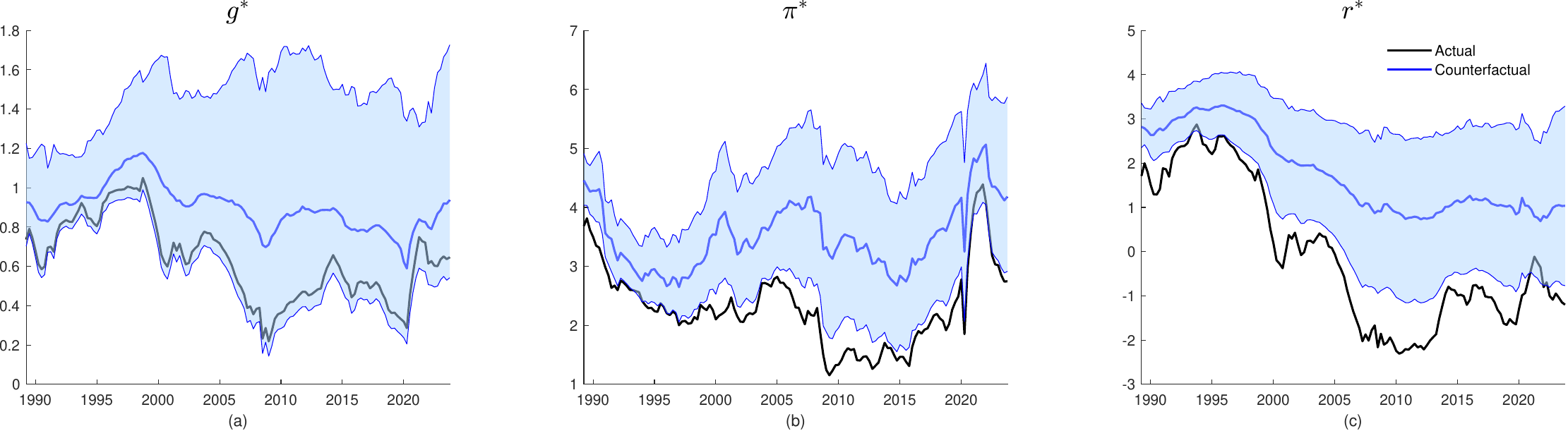}
	\caption{The plot compares the actual trends and the counterfactual path with 68\% credible bands. The counterfactual trends are calculated by shutting down the effects of all monetary policy shocks as is done in the calculation of historical decompositions, see \cite{kilian2017structural}. In this plot we estimate the shrinkage parameter $\lambda_B$ using a hierarchical Bayes approach. }
	\label{fig:Historical_estimatekappa}
\end{figure}

\begin{figure}[H]
	\centering
	\includegraphics[width=1\linewidth]{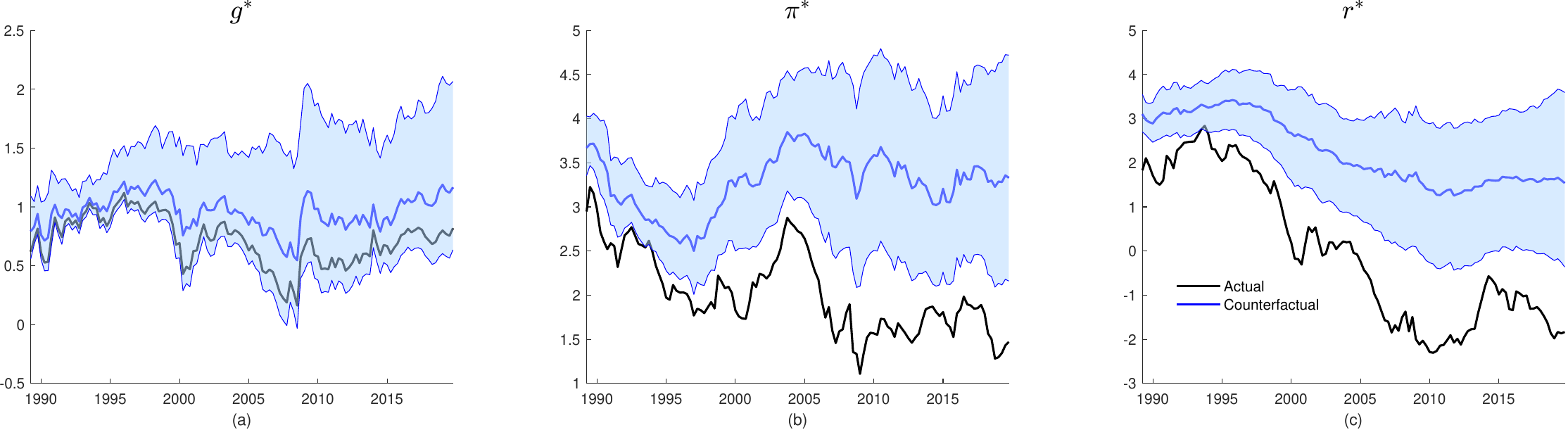}
	\caption{The plot compares the actual trends and the counterfactual path with 68\% credible bands. The counterfactual trends are calculated by shutting down the effects of all monetary policy shocks as is done in the calculation of historical decompositions, see \cite{kilian2017structural}. In this plot we use only data until 2019Q4. }
	\label{fig:Historical_excludeCovid}
\end{figure}

\begin{figure}[H]
	\centering
	\includegraphics[width=1\linewidth]{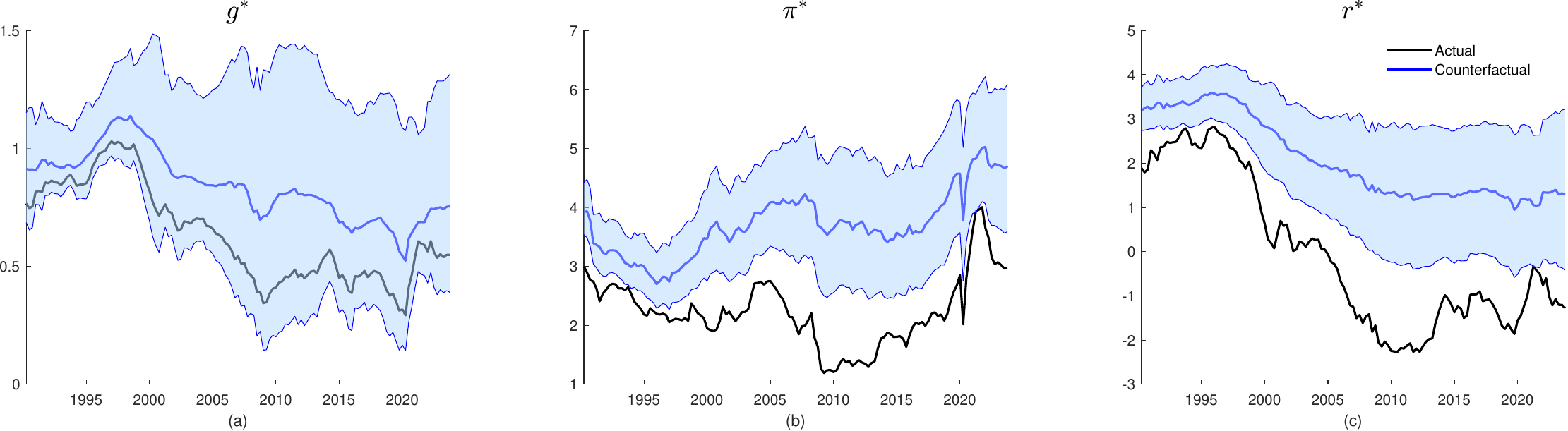}
	\caption{The plot compares the actual trends and the counterfactual path with 68\% credible bands. The counterfactual trends are calculated by shutting down the effects of all monetary policy shocks as is done in the calculation of historical decompositions, see \cite{kilian2017structural}. In this plot we estimate the model with eight lags instead of four. }
	\label{fig:Historical_P=8}
\end{figure}

\begin{figure}[H]
	\centering
	\includegraphics[width=1\linewidth]{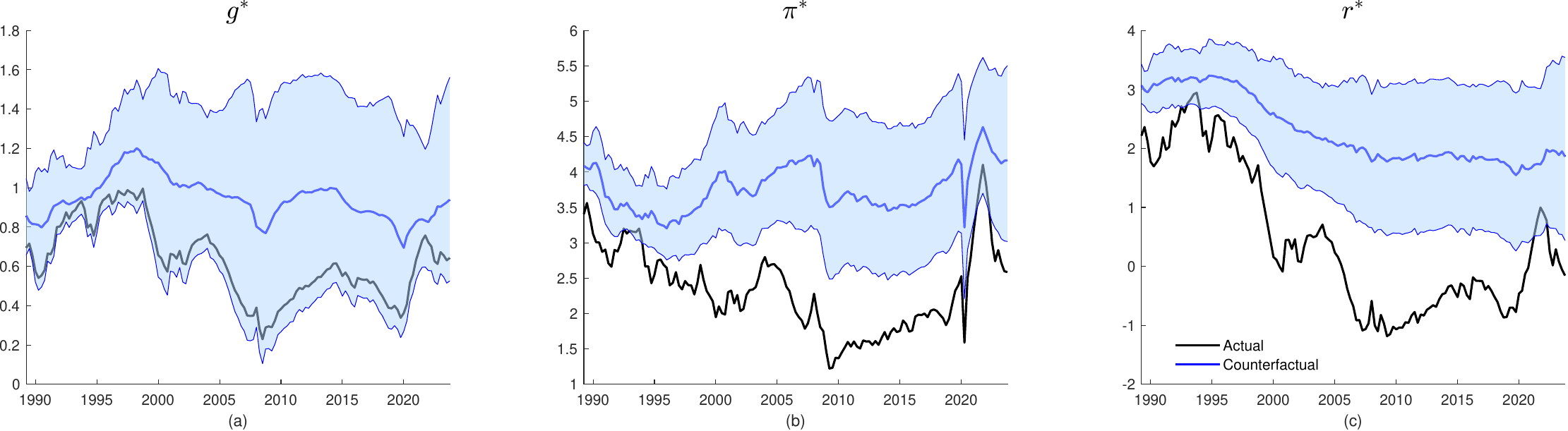}
	\caption{The plot compares the actual trends and the counterfactual path with 68\% credible bands. The counterfactual trends are calculated by shutting down the effects of all monetary policy shocks as is done in the calculation of historical decompositions, see \cite{kilian2017structural}. In this plot we estimate the model with the one-year treasury yield instead of using the federal funds effective rate. }
	\label{fig:Historical_GS1}
\end{figure}

\begin{figure}[H]
	\centering
	\includegraphics[width=1\linewidth]{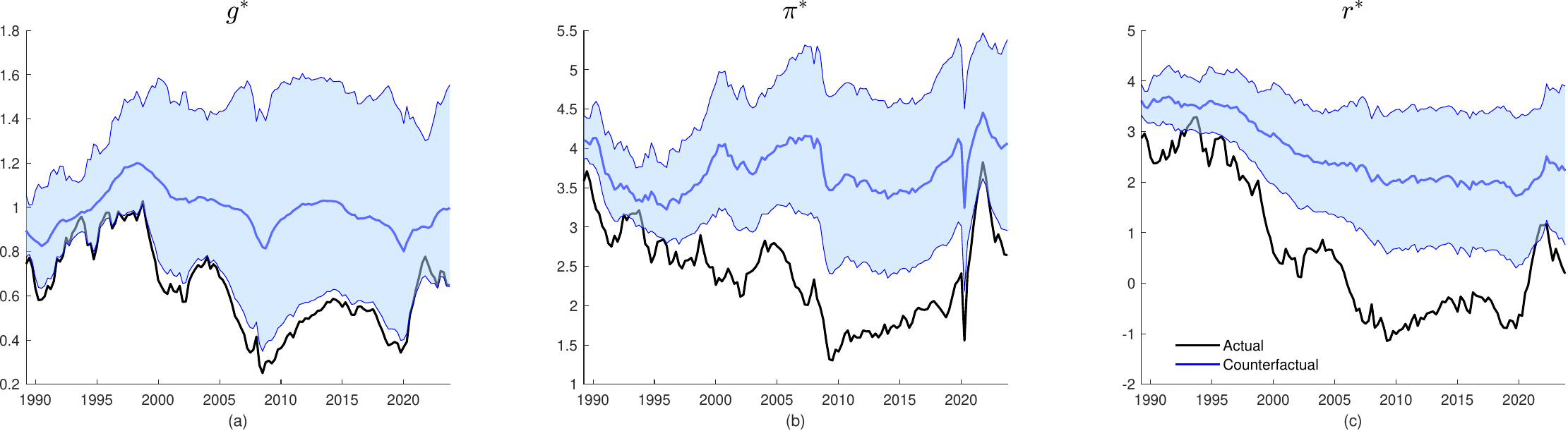}
	\caption{The plot compares the actual trends and the counterfactual path with 68\% credible bands. The counterfactual trends are calculated by shutting down the effects of all monetary policy shocks as is done in the calculation of historical decompositions, see \cite{kilian2017structural}. In this plot we estimate the model with the two-year treasury yield instead of using the federal funds effective rate. }
	\label{fig:Historical_GS2}
\end{figure}

\begin{figure}[H]
	\centering
	\includegraphics[width=1\linewidth]{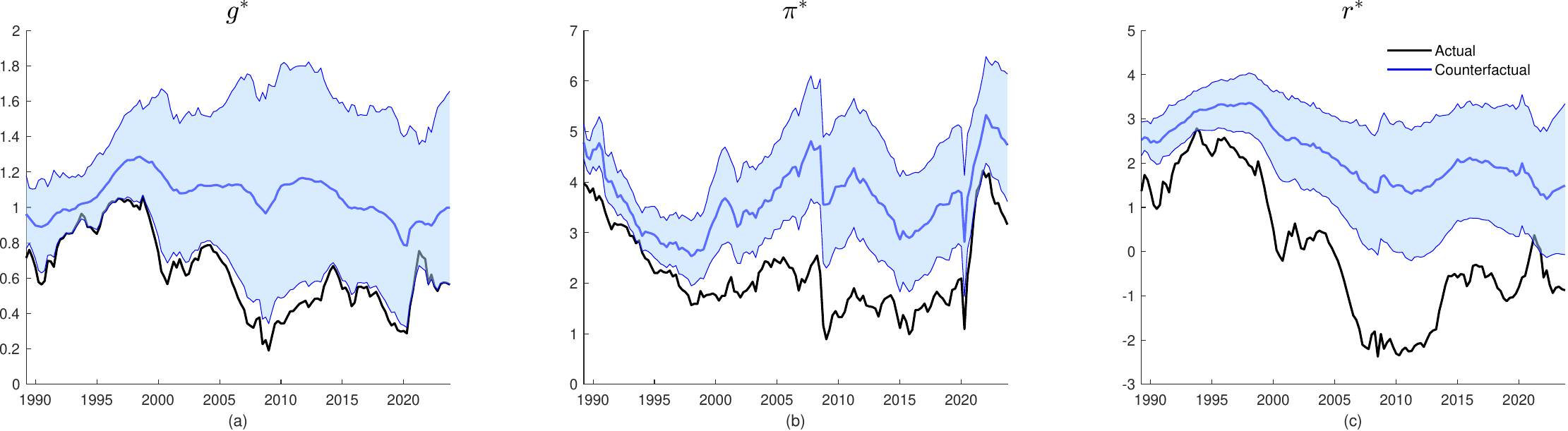}
	\caption{The plot compares the actual trends and the counterfactual path with 68\% credible bands. The counterfactual trends are calculated by shutting down the effects of all monetary policy shocks as is done in the calculation of historical decompositions, see \cite{kilian2017structural}. In this plot we estimate the model with the personal consumption expenditure price index instead of the GDP deflator. }
	\label{fig:Historical_PCEPI}
\end{figure}

\begin{figure}[H]
	\centering
	\includegraphics[width=1\linewidth]{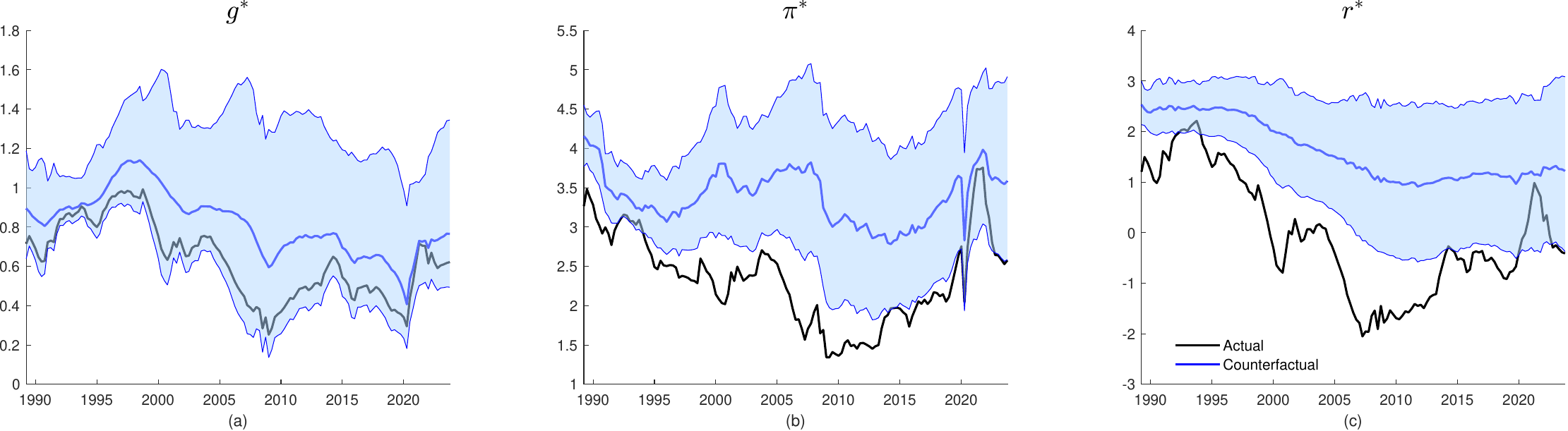}
	\caption{The plot compares the actual trends and the counterfactual path with 68\% credible bands. The counterfactual trends are calculated by shutting down the effects of all monetary policy shocks as is done in the calculation of historical decompositions, see \cite{kilian2017structural}. In this plot we estimate the model with unadjusted monetary policy surprise (MPS) measure from \citep{bauer2023reassessment} instead of using the orthogonalized monetary policy surprise measure. }
	\label{fig:Historical_MPS}
\end{figure}

\begin{figure}[H]
	\centering
	\includegraphics[width=1\linewidth]{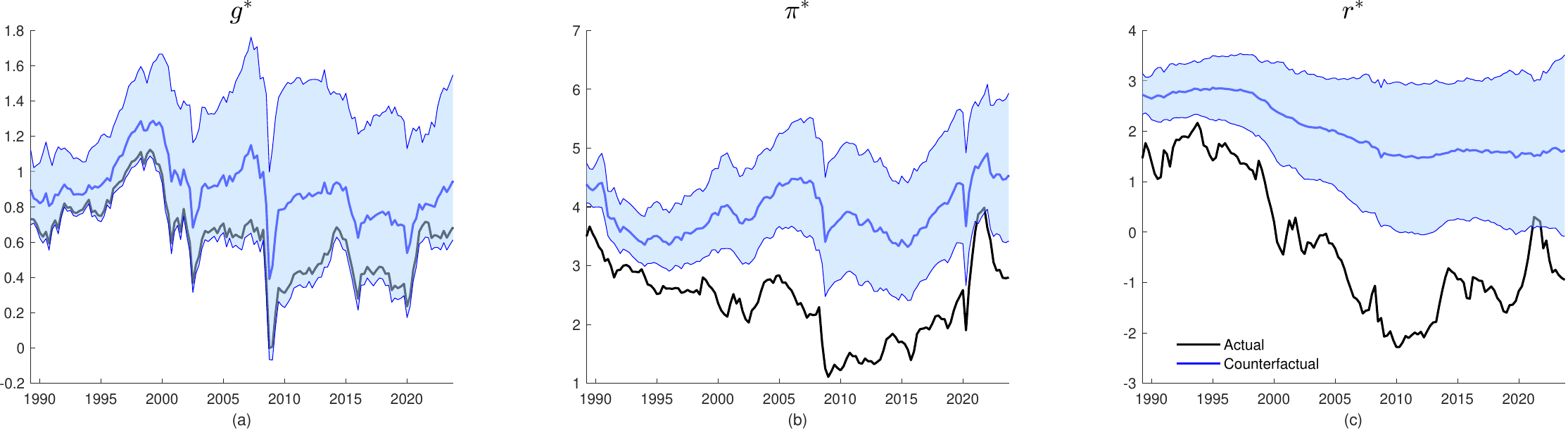}
	\caption{The plot compares the actual trends and the counterfactual path with 68\% credible bands. The counterfactual trends are calculated by shutting down the effects of all monetary policy shocks as is done in the calculation of historical decompositions, see \cite{kilian2017structural}. In this plot we estimate the model with adding the excess bond premium.   }
	\label{fig:Historical_EBP}
\end{figure}

	\end{document}